\documentclass[apj]{emulateapj}
\usepackage{amsmath}
\usepackage{amssymb}
\usepackage{mathrsfs}
\usepackage{natbib}
\pdfoutput=1

\shorttitle{NIR H$_{2}$O and OH in HD~101412}
\shortauthors{Adams et al.}

\begin{document}

\title{Water and OH Emission from the inner disk of a Herbig Ae/Be star}

\author{Steven C. Adams}
\affil{Department of Physics and Astronomy, 118 Kinard Laboratory, Clemson University, Clemson, SC 29634-0978}
\author{M\'at\'e \'Ad\'amkovics}
\affil{Department of Physics and Astronomy, 118 Kinard Laboratory, Clemson University, Clemson, SC 29634-0978}
\author{John S. Carr}
\affil{Naval Research Laboratory, Code 7211, Washington, D.C. 20375}
\author{Joan R. Najita}
\affil{National Optical Astronomy Observatory, 950 North Cherry Avenue, Tucson, AZ 85719}
\author{Sean D. Brittain}
\affil{Department of Physics and Astronomy, 118 Kinard Laboratory, Clemson University, Clemson, SC 29634-0978 \\
National Optical Astronomy Observatory, 950 North Cherry Avenue, Tucson, AZ 85719; sbritt@clemson.edu}

\begin{abstract}

We report the detection of hot H$_{2}$O  and OH emission from the Herbig Ae/Be star HD~101412 using the Cryogenic Infrared Echelle Spectrograph on the \textit{Very Large Telescope}. 
Previous studies of Herbig Ae/Be stars have shown the presence of OH around some of these sources, but H$_{2}$O has proven more elusive. While marginal water emission has been 
reported in the mid-infrared, and a few Herbig Ae/Be stars show water emission in the far-infrared, water emission near 2.9 $\mu$m has not been previously detected. 
We apply slab models to the ro-vibrational OH, H$_{2}$O, and CO spectra of this source and show that the molecules are consistent with being cospatial. We discuss 
the possibility that the detection of the CO overtone bandhead emission, detection of water emission, and the large line to continuum contrast of the OH lines may be 
connected to its high inclination and the $\lambda$ Bo\"o nature of this star. If the low abundance of refractories results from the selective accretion of gas relative to dust, the 
inner disk of HD~101412 should be strongly dust-depleted allowing us to probe deeper columns of molecular gas in the disk, enhancing its molecular emission. Our detection 
of C- and O-bearing molecules from the inner disk of HD~101412 is consistent with the expected presence in this scenario of abundant volatiles in the accreting gas. 

\end{abstract}


\section{Introduction}

High resolution spectroscopic studies of Herbig Ae/Be (HAeBe) stars indicate that their circumstellar environments are commonly home to hot CO 
\citep{Blake04, Brittain07, Salyk11, Salyk11_2, Brown13, vanderPlas15, Banzatti15} and, less frequently, OH gas \citep{Mandell08,Fedele11,Brittain16},
but they have yet to yield a detection of the H$_{2}$O emission from 2-4 $\mu$m that has been observed in lower-mass pre-main sequence T~Tauri systems 
\citep{Carr04,Salyk08,Fedele11,Doppmann11,Mandell12,Banzatti17}. 

Likewise, mid- and far-infrared studies have yielded few detections of H$_{2}$O emission from cooler gas in disks around HAeBes. In a study of 25 HAeBes using \textit{Spitzer}, there 
were marginal detections of H$_{2}$O emission reported for 12 of 25 systems \citep{Pontoppidan10}. These detections were reported based on inspection by eye and were not 
determined to be above the 3.5$\sigma$ detection threshold defined in the study. In contrast, similar surveys of T Tauri stars yield a much higher detection rate for H$_{2}$O 
emission in the mid-infrared (22 of 48 exhibited a significant detection of H$_{2}$O emission; \citealt{Pontoppidan10}). Statistics compiled by \citet{Banzatti17} showed 
that between 63-85 \% out of 64 stars with a stellar mass less than 1.5 M$_{\odot}$ exhibit mid-infrared H$_{2}$O emission. In searches for H$_{2}$O in the 
far-infrared, only 3 sources were detected out of 25 HAeBes observed \citep{Meeus12,Fedele12,Fedele13}. These results suggests that the abundance of H$_{2}$O 
gas in the optically thin upper atmosphere around HAeBes is low. 

One possible explanation for the dearth of near-infrared (NIR) water detections 
among HAeBes compared to T Tauri stars is that the high far-ultraviolet 
(FUV) luminosity of HAeBe stars causes a 
larger column of water to be dissociated to produce OH (e.g., \citealt{Adamkovics16,Najita17}). 
As the water falls below the dust photosphere, emission from the water becomes 
impossible to detect. Here we report what appears to be an exception to this picture - the first detection of NIR H$_{2}$O emission from the HAeBe star HD~101412.

HD~101412 is a B9.5Ve star \citep{Valenti03} located at a distance of 411$^{+5}_{-4}$ pc \citep{Gaia16,Gaia18}. As part of an X-Shooter survey of 92 HAeBes, 
\cite{Fairlamb15} determined the stellar parameters of their sample self-consistently. They find that T$_{\rm{eff}}$ = 9750 $\pm$ 250 K and log(L/L$_{\odot}$) = 1.36 $\pm$ 0.23
adopting d = 301 pc. We adopt the Gaia distance and use Siess pre-main sequence models to recalculate the stellar mass, radius, and luminosity \citep{Siess00}. The updated 
values are M = 2.5 M$_{\odot}$, R = 2.3 R$_{\odot}$, and log(L/L$_{\odot}$) = 1.63. 

The inner disk surrounding HD~101412 is nearly edge-on. Fitting a uniform ring model to $N-$band visibilities acquired with MIDI on the VLT indicates that the disk is inclined 
80$\degr$ $\pm$ 7$\degr$ \citep{Fedele08}. No (sub)mm observations have been made of this source so there is no observational estimate of the extent or mass of the disk. 
The mid-infrared SED of the star indicates that it is a self-shadowed disk (Group II; \citealt{Fedele08}). \citet{Fairlamb17} report the flux of H$\alpha$ for this source and provide 
a relationship between the accretion luminosity and luminosity of the H$\alpha$ line.  Adopting the stellar parameters above, we find that the accretion rate is 1.6 $\times$ 
10$^{-7}$ M$_{\sun}$ yr$^{-1}$. The accretion rate indicates that HD~101412 still harbors a large gaseous reservoir.

The  disk of HD~101412 reveals a rich molecular spectrum. Both the ro-vibrational CO overtone \citep{Cowley12,Ilee14,vanderPlas15} and fundamental \citep{vanderPlas15} 
emission lines have been observed. Modeling of the profile of these lines indicate that the emitting region is narrow (0.8 - 1.2 AU; \citealt{vanderPlas15}). Mid-infrared 
spectroscopy also reveals CO$_{2}$ emission from the disk of HD~101412 \citep{Pontoppidan10,Salyk11}. CO$_{2}$ emission and CO bandheads of the first overtone emission are 
both unusual features to observe in the spectrum of HAeBe stars. Only 7\% show CO overtone bandhead emission in a survey of 91 HAeBes \citep{Ilee14} and only 4\% show 
CO$_{2}$ emission in a survey of 25 HAeBes \citep{Salyk11}. Here we report the detection of ro-vibrational OH and H$_{2}$O emission from this source as well (Sections 2 and 3). 
We apply slab models of CO, OH, and H$_{2}$O to compare the emitting radii, temperature, and column densities of these molecules (Section 4), and compare the molecular emission 
observed in HD~101412 to other young stellar objects (Section 5). Finally, we discuss the implications for our understanding of the molecular content of the inner disks around 
HAeBe stars (Section 6).

\section{Observations}

$L$-band observations of HD~101412 were obtained from the European Southern Observatory (ESO) Data Archives along with the associated raw calibration files, based on observations 
under programme ID 091.C-0796(A). The data were acquired on 27 May 2013 using a 0.$^{\prime\prime}$2 slit width at a central wavelength of 2.94 $\mu$m using the Cryogenic Infrared 
Echelle Spectrograph (CRIRES; \citealt{Kaufl04}) on the ESO \textit{Very Large Telescope} (\textit{VLT}) UT1. CRIRES has four detectors that each cover $\sim$0.0160 $\mu$m with a 
$\sim$0.0045 $\mu$m gap between each chip. Exposures were taken in an \textit{ABBA} nod pattern with a 10.$^{\prime\prime}$ nod in order to remove sky emission lines. An integration 
time of 60 s and 3 sub-integrations were used for each of 20 exposures giving a total integration time of 3600 s. Table \ref{tab:Obs} gives details of all observations used in this study.

\begin{deluxetable}{lcccc}
\tabletypesize{\scriptsize}
\tablecaption{Observations}
\tablewidth{0pt}
\tablehead{
\colhead{Star} & \colhead{Date} & \colhead{Airmass} & \colhead{Exposures} & \colhead{Int. Time (s)}
}
\startdata
\multicolumn{5}{c}{$L-$Band} \\
\hline
HD~101412 & 20130527 & 1.265 & 20 & 3600 \\
$\lambda$ Cen & 20130527 & 1.277 & 8 & 240 \\
\hline
\multicolumn{5}{c}{$K-$Band} \\
\hline
HD~101412 & 20110405 & 1.250 & 8 & 4800 \\
j Cen & 20110405 & 1.301 & 8 & 640
\enddata
\label{tab:Obs}
\tablecomments{
Observation information for HD~101412 OH, H$_{2}$O, and CO data presented. All data were obtained using the ESO Data Archives.
}
\end{deluxetable}

\begin{figure*}[t]
\includegraphics[width=7.0in,height=6.5in,trim={1.5in 0.7in 1.5in 0.5in},clip]{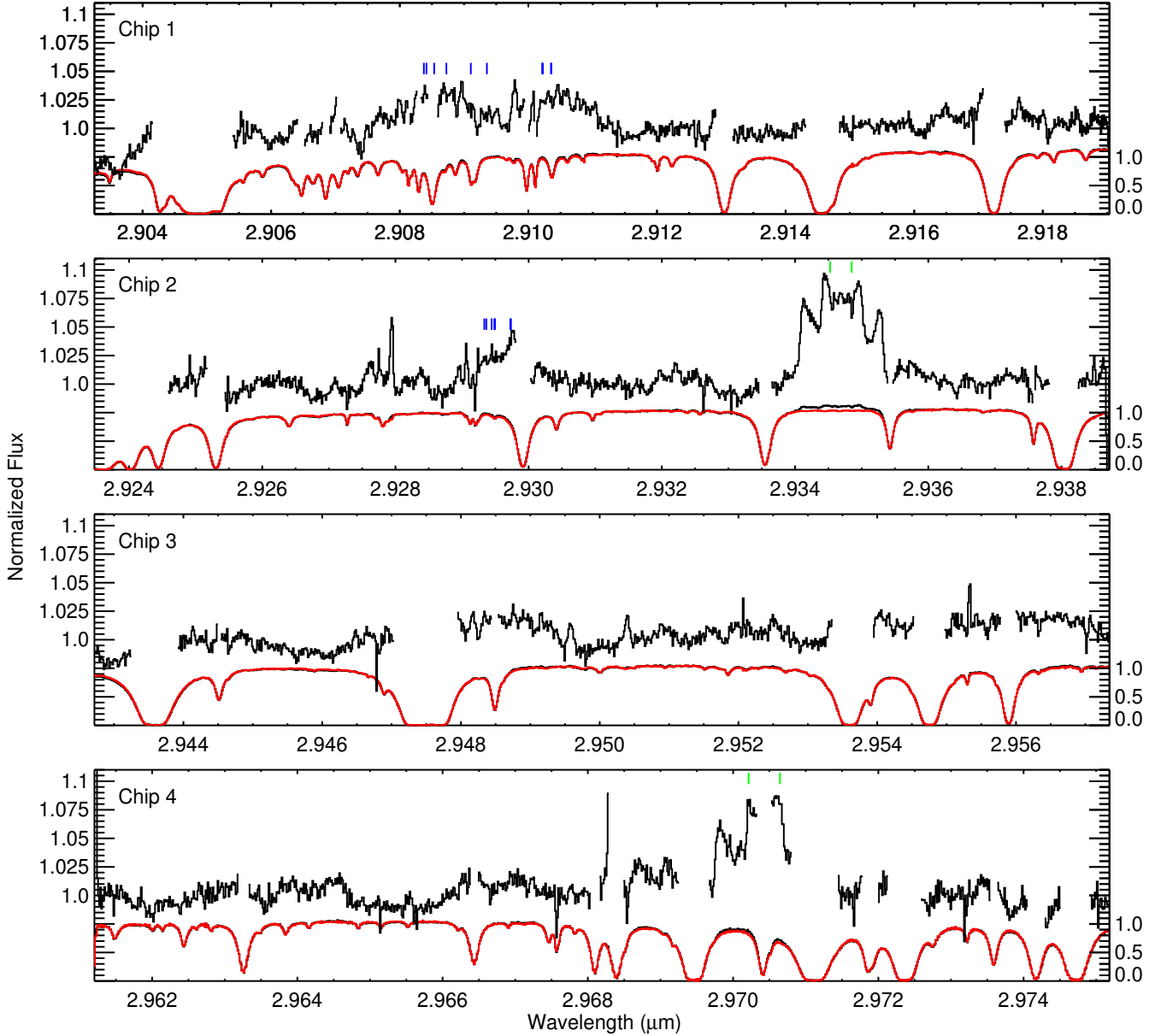}
\caption{
$L-$band observations of HD~101412. Chip 1 is the upper plot and chip 4 is the lower plot, with the gap between each plot resulting from the gaps between the chips. In 
each panel we plot the normalized spectrum of HD~101412 (black) and the telluric standard $\lambda$ Cen (HR 4467) (red). The telluric corrected spectrum of HD~101412 is 
plotted above (stretched by a factor of 20 and offset by 0.5 units). The left axis shows the normalized flux units scaled to the ratioed spectrum while the right axis shows 
the normalized flux units for the science and standard star spectra. The locations of the OH emission doublets are marked with green hashes with the structure 
observed in the continuum coming from H$_{2}$O emission, with the strongest H$_{2}$O features marked with blue hashes.
}
\label{fig:OH}
\end{figure*}

Data reduction was performed using software based on algorithms developed for the reduction of PHOENIX and NIRSPEC data (described in \citealt{Brittain07}). Flats 
and darks were taken in order to remove systematic variation in pixel gain. Sequential \textit{AB} observations were combined (\textit{A} - \textit{B}) and 
then divided by the normalized flat field image. Median values of the combined images were used to identify and remove hot and bad pixels, as well as cosmic ray hits. 
Pixel values that differ by 6$\sigma$ were rejected. Spectra were then extracted using a rectangular extraction method. Wavelength calibration was performed 
using the telluric absorption features observed in the spectrum. A Sky Synthesis Program (SSP) model atmosphere \citep{Kunde74}, which accesses the 2003 
HITRAN molecular database \citep{Rothman03}, was computed based on the airmass of the observations. Standard star observations of $\lambda$ Cen were taken 
immediately after the observations of HD~101412. The telluric standard was reduced following the same process as HD~101412. The normalized spectrum of HD~101412 
was divided by the normalized spectrum of $\lambda$ Cen to correct for atmospheric absorption lines. Regions of the spectrum where the atmospheric transmittance 
was below 50\% were excluded. Final reduced $L-$band spectra and ratios are presented in Figure \ref{fig:OH}.

We also reduced archival data for the CO bandhead emission previously reported for HD~101412 \citep{Cowley12,Ilee14}. The data were obtained based on observations 
made with CRIRES on the \textit{VLT} under programme ID 087.C-0124(A). The CO data were re-reduced using the same method as the OH and H$_{2}$O observations. 
The CO data were then modeled in order to self-consistently determine the CO emitting region, temperature, and column density.

The flux densities adopted for the continua of the $K$- and $L$-band spectra were obtained using values from Johnson:K and Johnson:L filter photometry measurements 
found using the VizieR Photometry Viewer \citep{Ochsenbein00}. The flux density at 2.94 $\mu$m was estimated at a value between the flux density at these filters by a 
linear fit to the two data points. A continuum flux of 3.20 $\times$ 10$^{-10}$ erg s$^{-1}$ cm$^{-2}$ $\mu$m$^{-1}$ was used for the $L$-band (OH) and 4.21 $\times$ 10$^{-10}$ 
erg s$^{-1}$ cm$^{-2}$ $\mu$m$^{-1}$ was used for the $K$-band (CO).

\section{Results}

The fully reduced $L-$band spectrum has a spectral resolution of R $\approx$ 90,000 and a signal to noise ratio of $\sim$200. We detect the OH P4.5 and OH P5.5 doublets 
with the peak in the normalized line flux relative to the continuum of 9\% (Figure \ref{fig:ohall}, panel A and C, respectively). Gaps in the profile of the P5.5 
doublet are due to telluric absorption greater than 50\%. We also detect H$_{2}$O emission near 2.93 $\mu$m (Figure \ref{fig:h2o_2}). Another H$_{2}$O feature between 
2.9074 and 2.9110 $\mu$m is also observed (Figure \ref{fig:h2o_1}). Table \ref{tab:MolLines} gives transition parameters for the individual transitions that we propose 
comprise the most prominent emission features. Errors for equivalent widths are determined by adding the noise across each pixel in the emission feature in 
quadrature.

\begin{deluxetable}{lcccc}
\tabletypesize{\scriptsize}
\tablecaption{Molecular Line Properties}
\tablewidth{0pt}
\tablehead{
\colhead{Wavelength} & \colhead{Transition} & \colhead{A$_{ul}$} & \colhead{g$_{u}$} & \colhead{E$_{u}$} \\
\colhead{($\mu$m)} & \colhead{} & \colhead{(s$^{-1}$)} & \colhead{} & \colhead{(K)}
}
\startdata
\multicolumn{5}{c}{OH Lines} \\
\hline
2.97040 & 3/2 P5.5f & 12.67 & 20 & 5627.64 \\
2.96997 & 3/2 P5.5e & 12.67 & 20 & 5626.56 \\
\\
2.93461 & 3/2 P4.5f & 11.68 & 16 & 5414.92 \\
2.93428 & 3/2 P4.5e & 11.68 & 16 & 5414.31 \\
\\
\hline
\multicolumn{5}{c}{H$_{2}$O Lines} \\
\hline 
2.92949 & 10 $_{8}$ $_{2}$ $\rightarrow$ 11 $_{8}$ $_{3}$ & 18.45 & 63 & 8540.39 \\
2.92948 & 10 $_{8}$ $_{3}$ $\rightarrow$ 11 $_{8}$ $_{4}$ & 21.20 & 21 & 8540.39 \\
2.92925 & 11 $_{5}$ $_{6}$ $\rightarrow$ 12 $_{5}$ $_{7}$ & 39.56 & 23 & 8221.99 \\
2.92924 & 14 $_{2}$ $_{13}$ $\rightarrow$ 15 $_{2}$ $_{14}$ & 48.79 & 29 & 8697.67 \\
2.92920 & 14 $_{1}$ $_{13}$ $\rightarrow$ 15 $_{1}$ $_{14}$ & 48.58 & 87 & 8697.71 \\
2.92912 & 13 $_{3}$ $_{11}$ $\rightarrow$ 14 $_{3}$ $_{12}$ & 48.45 & 81 & 8583.04 \\
2.92909 & 13 $_{2}$ $_{11}$ $\rightarrow$ 14 $_{2}$ $_{12}$ & 48.94 & 27 & 8582.23 \\
\\
2.91011 & 13 $_{2}$ $_{12}$ $\rightarrow$ 14 $_{2}$ $_{13}$ & 48.78 & 81 & 8293.46 \\
2.91011 & 13 $_{1}$ $_{11}$ $\rightarrow$ 14 $_{1}$ $_{12}$ & 48.91 & 27 & 8293.43 \\
2.90998 & 12 $_{2}$ $_{10}$ $\rightarrow$ 13 $_{2}$ $_{11}$ & 49.17 & 75 & 8177.10 \\
2.90997 & 12 $_{3}$ $_{10}$ $\rightarrow$ 13 $_{3}$ $_{11}$ & 48.16 & 25 & 8178.82 \\
2.90911 & 10 $_{6}$ $_{4}$ $\rightarrow$ 11 $_{6}$ $_{5}$ & 31.81 & 63 & 8030.62 \\
2.90886 & 11 $_{3}$ $_{8}$ $\rightarrow$ 12 $_{3}$ $_{9}$ & 53.09 & 23 & 7976.11 \\
2.90848 & 10 $_{6}$ $_{5}$ $\rightarrow$ 11 $_{6}$ $_{6}$ & 31.57 & 21 & 8029.60 \\
2.90829 & 11 $_{4}$ $_{8}$ $\rightarrow$ 12 $_{4}$ $_{9}$ & 42.70 & 69 & 8004.54 \\
2.90817 & 14 $_{1}$ $_{14}$ $\rightarrow$ 15 $_{1}$ $_{15}$ & 48.52 & 29 & 8340.48 \\
2.90813 & 14 $_{0}$ $_{14}$ $\rightarrow$ 15 $_{0}$ $_{15}$ & 48.04 & 81 & 8340.56 \\
\enddata
\label{tab:MolLines}
\tablecomments{
Parameters for some transitions that comprise observed emission features in HD~101412 $L-$band observations. Line groups are presented in the order that the 
emission feature is discussed in the text. All data were acquired using the HITRAN database \citep{Rothman13}.
}
\end{deluxetable}

\subsection{OH Emission}

\begin{figure*}[t]
\includegraphics[width=7.in,height=2.8in,trim={0.0in 2.in 0.0in 2.0in},clip]{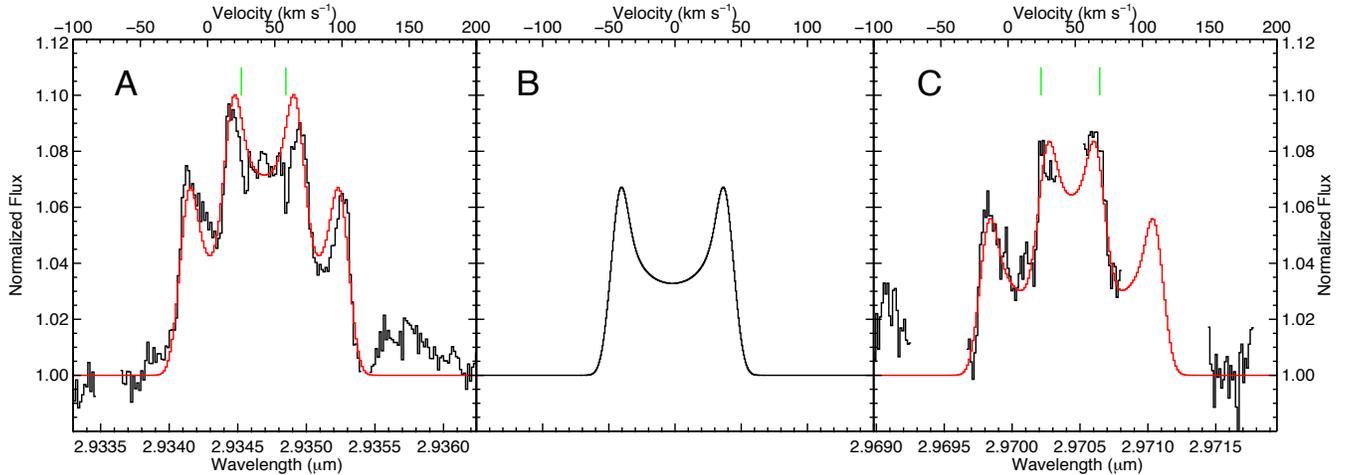}
\caption{
(A) P4.5 (1+, 1$-$) OH emission doublet. The normalized fluxes of the OH doublets are plotted versus wavelength. The upper x-axis shows the relative velocity of the emission 
features. Zero velocity is centered at the laboratory rest wavelength of the P4.5 (1+) line (2.93428 $\mu$m). The green tick marks indicate the wavelength of each doublet 
feature in the rest frame of the star. The Doppler shift inferred from the molecular emission indicates that the heliocentric radial velocity is 16.9 km s$^{-1}$, which is consistent with 
the heliocentric radial velocity inferred from the measurement of photospheric lines (16.5 km s$^{-1}$; \citealt{Hubrig10}).  The best fit model is plotted in red. 
(B) Model line profile of each doublet feature, scaled to match P4.5 intensity. (C) P5.5 (1+, 1$-$) OH  emission feature. The emission feature is partially obscured by telluric 
absorption. The green tick marks indicate the location of each doublet feature in the rest frame of the star. The zero velocity bin is centered at 2.96997 $\mu$m, which is the 
laboratory rest wavelength of the P5.5 (1+) lines.
}
\label{fig:ohall}
\end{figure*}

The P4.5 doublet transition is spectrally resolved, however, the doublet itself is blended. The emission feature is bracketed by strong telluric absorption features. The 
equivalent width (EW) is calculated over the entire range of the doublet between the absorption features and divided by 2 due to the blending. The EW is 4.3 $\pm$ 0.2 $\times$ 
10$^{-5}$ $\mu$m which, when factoring in the distance and $L$-band flux density, corresponds to a line luminosity of 7.2 $\pm$ 0.3 $\times$ 10$^{-5}$ L$_{\odot}$. 
The line to continuum contrast of the P4.5 OH doublet is 9\% (Figure \ref{fig:ohall}) which is more than three times the line to continuum contrast typically observed 
for this doublet in previous observations of HAeBes \citep{Mandell08,Fedele11,Brittain16}.

The P5.5 emission feature is partially obscured by atmospheric absorption. The profile of the P5.5 doublet and P4.5 doublet differ slightly in the blue portion due to 
different separations of the doublet transition energies. Thus, the individual peaks in the blue portion of the P5.5 feature show each doublet's peak as being further 
apart. This broadens the overall line width, however, the inner peak separation between the blue portion of the (1$-$) line and red portion of the (1+) line is reduced. 
Due to the atmospheric absorption, an EW cannot be determined from the data. Based on the model fits obtained from fitting the full P4.5 doublet, we determine the P5.5 
EW to be 3.7 $\pm$ 0.1 $\times$ 10$^{-5}$ $\mu$m which corresponds to a line luminosity of 6.2 $\pm$ 0.3 $\times$ 10$^{-5}$ L$_{\odot}$.

We observe a Doppler shift of 24.4 km s$^{-1}$. This implies a heliocentric radial velocity of 16.9 km s$^{-1}$ based on the date the observations were made. \cite{Hubrig10} observe Fe 
lines in the spectrum of HD~101412 and report an average heliocentric radial velocity of 16.65 km s$^{-1}$, which is consistent with our determined radial velocity.

\begin{figure}
\includegraphics[width=3.15in, height=3.in,keepaspectratio,trim={2.in 1.in 3.in 1.5in},clip]{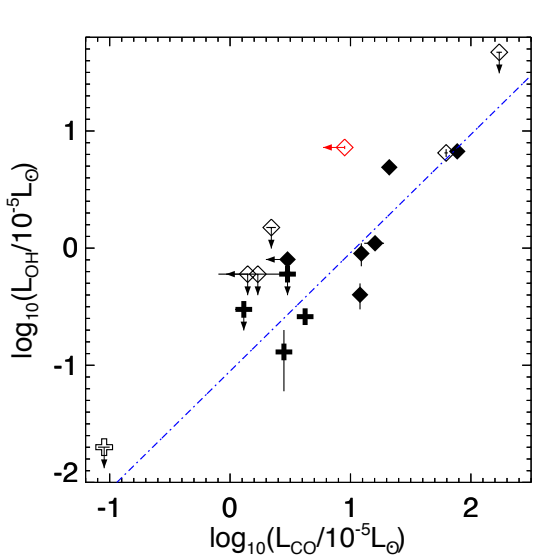}
\caption{
Adapted from \citet{Brittain16}. HD~101412 has been added and is plotted in red. Group I HAeBe stars are marked as filled symbols. Group II HAeBe stars are marked as open 
symbols. The disks that have been observed to have an optically thin inner disk are labeled with ``plus" signs. Disks without such optically thin regions are labeled with 
diamonds. The dotted-dashed line is a linear least-squares fit to the detections, not including HD~101412.
}
\label{fig:ohco}
\end{figure}

\cite{Brittain16} find a power-law relationship between the luminosity of ro-vibrational CO ($\nu$ = 1 $\rightarrow$ 0 P30) and OH ($\nu$ = 1 $\rightarrow$ 0 P4.5) 
emission from HAeBes and find that the ratio of their luminosities is 11.0$\pm$0.2. We compare the relative luminosity of the OH and CO emission for HD~101412  (Figure \ref{fig:ohco}). 
Because the lines are so broad, there is significant line blending. We take the most isolated CO line ($\nu$ = 1 $\rightarrow$ 0 P26 transition; \citealt{Troutman10}) and determine the 
luminosity of the P30 line assuming the gas is 1300~K (see section 4). Because of the line blending, we take the CO luminosity to be an upper limit and find that L(CO)/L(OH) $\leq$ 
1.24. Thus the relative flux of the OH emission is an order of magnitude larger than the previous HAeBes studied.

\subsection{H$_{2}$O Emission}

\begin{figure}
\includegraphics[width=3.5in,height=2.5in,trim={0.6in 0.6in 0.7in 0.55in},clip]{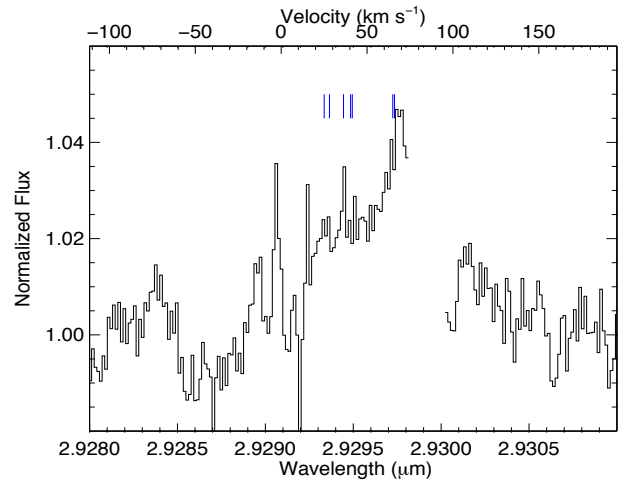}
\caption{
The 2.93 $\mu$m H$_{2}$O emission lines. The blue tick marks indicate the positions of the dominant transitions contributing to the emission feature. The zero velocity bin is 
centered at 2.92909 $\mu$m, the transition with the shortest wavelength of the list in Table \ref{tab:MolLines} found in this spectral region. The upper x-axis gives the velocity 
space information and shows a Doppler shift of 24.4 km s$^{-1}$, consistent with the OH lines. 
}
\label{fig:h2o_2}
\end{figure}

One prominent H$_{2}$O emission feature is observed at 2.929 $\mu$m (Figure \ref{fig:h2o_2}). This feature is partially obscured by telluric absorption. The emission is due 
to a blend of multiple transitions. Table \ref{tab:MolLines} presents some transitions that comprise the emission feature.

\begin{figure}
\includegraphics[width=3.5in,height=2.5in,trim={0.6in 0.6in 0.7in 0.55in},clip]{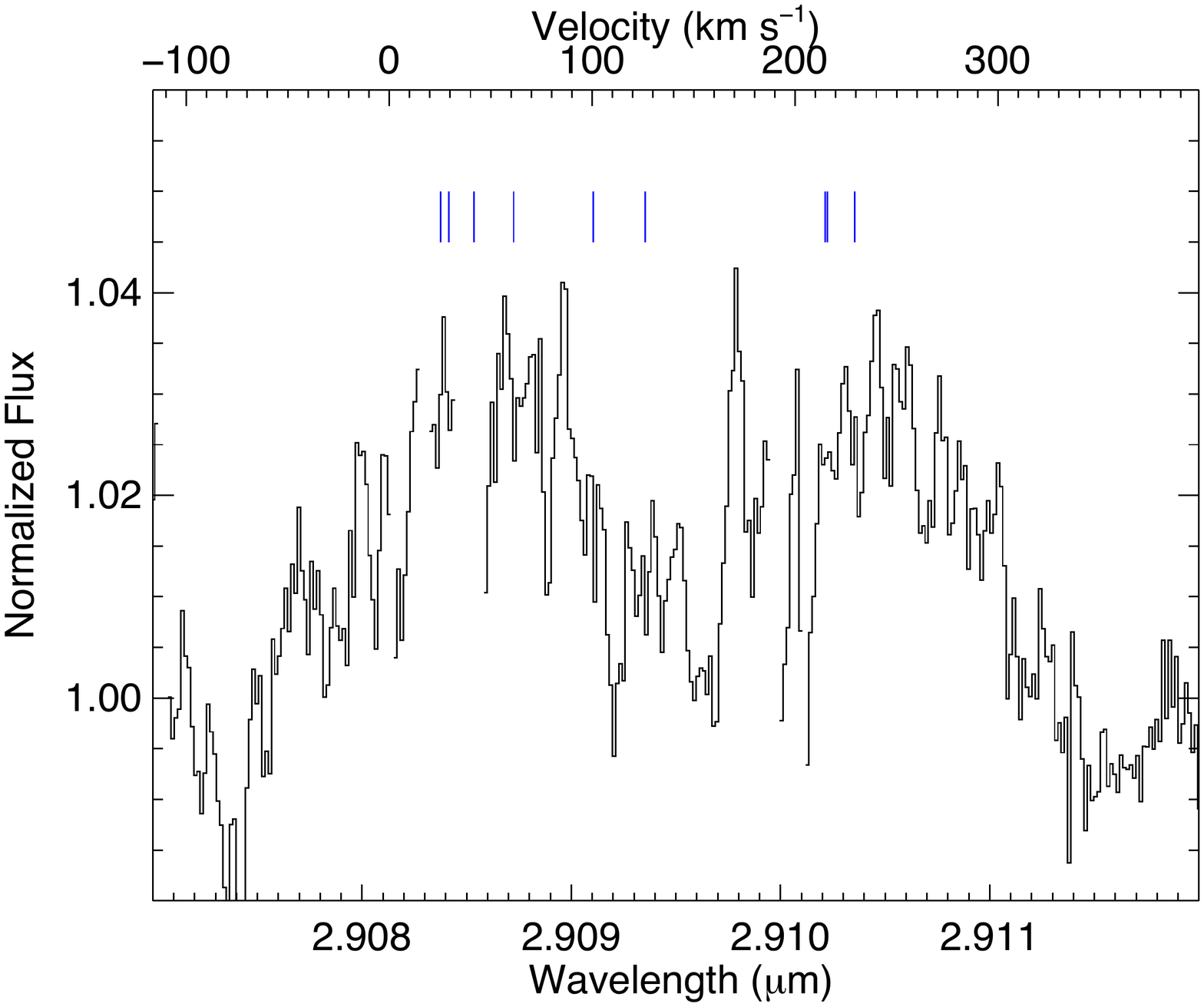}
\caption{
The 2.91 $\mu$m H$_{2}$O emission lines. The blue tick marks indicate the positions of the dominant transitions contributing to the emission feature. The zero velocity bin is centered at 
2.90813 $\mu$m, the transition with the shortest wavelength of the list in Table \ref{tab:MolLines} found in this spectral region. The upper x-axis gives the velocity space information 
and shows a Doppler shift of 24.4 km s$^{-1}$. H$_{2}$O transition information can be found in Table \ref{tab:MolLines}.
}
\label{fig:h2o_1}
\end{figure}

Another H$_{2}$O emission feature is observed between 2.9074 and 2.9110 $\mu$m (Figure \ref{fig:h2o_1}), with some regions obscured by atmospheric absorption. 
This feature is also a blend of multiple transitions. In both instances, the H$_{2}$O transitions observed all require high temperatures to reach the upper levels, 
thus making it unlikely that the H$_{2}$O emission observed is residual from telluric correction. 

\subsection{CO Observations}

In order to self-consistently determine the column densities of CO, OH, and H$_{2}$O, we also present $K-$band observations of HD~101412. We reproduce the results of \citet{Cowley12} 
and \citet{Ilee14} in that we detect both the CO $\nu$ = 2 $\rightarrow$ 0 and $\nu$ = 3 $\rightarrow$ 1 bandheads. We determine the signal-to-noise of the chips containing 
the CO $\nu$ = 2 $\rightarrow$ 0 bandhead and isolated emission features (Chips 2 and 3) to be $\sim$290, while Chip 4, containing the $\nu$ = 3 $\rightarrow$ 1 bandhead emission, has a 
signal-to-noise of $\sim$120. Figure \ref{fig:CO20iso} shows the isolated CO $\nu$ = 2 $\rightarrow$ 0 emission lines (Chip 3), while Figure \ref{fig:CObandmod}
show the CO $\nu$ = 2 $\rightarrow$ 0 (left) and $\nu$ = 3 $\rightarrow$ 1 (right) bandhead emission. Discussion of the modeling is presented in Section 4.

\section{Modeling}

To determine the spatial location, column density, and temperature of the CO, OH, and H$_{2}$O emission, we fit the spectra using a slab model \citep{Carr04}. The 
disk models assume LTE and Keplerian rotation. Because we find that the emission originates in a fairly narrow annulus, the gas temperature and the column density 
for each species are taken to be constant over the emitting region.

\begin{figure}
\includegraphics[width=3.5in,height=2.3in,trim={0.5in 1.0in 0.7in 1.2in},clip]{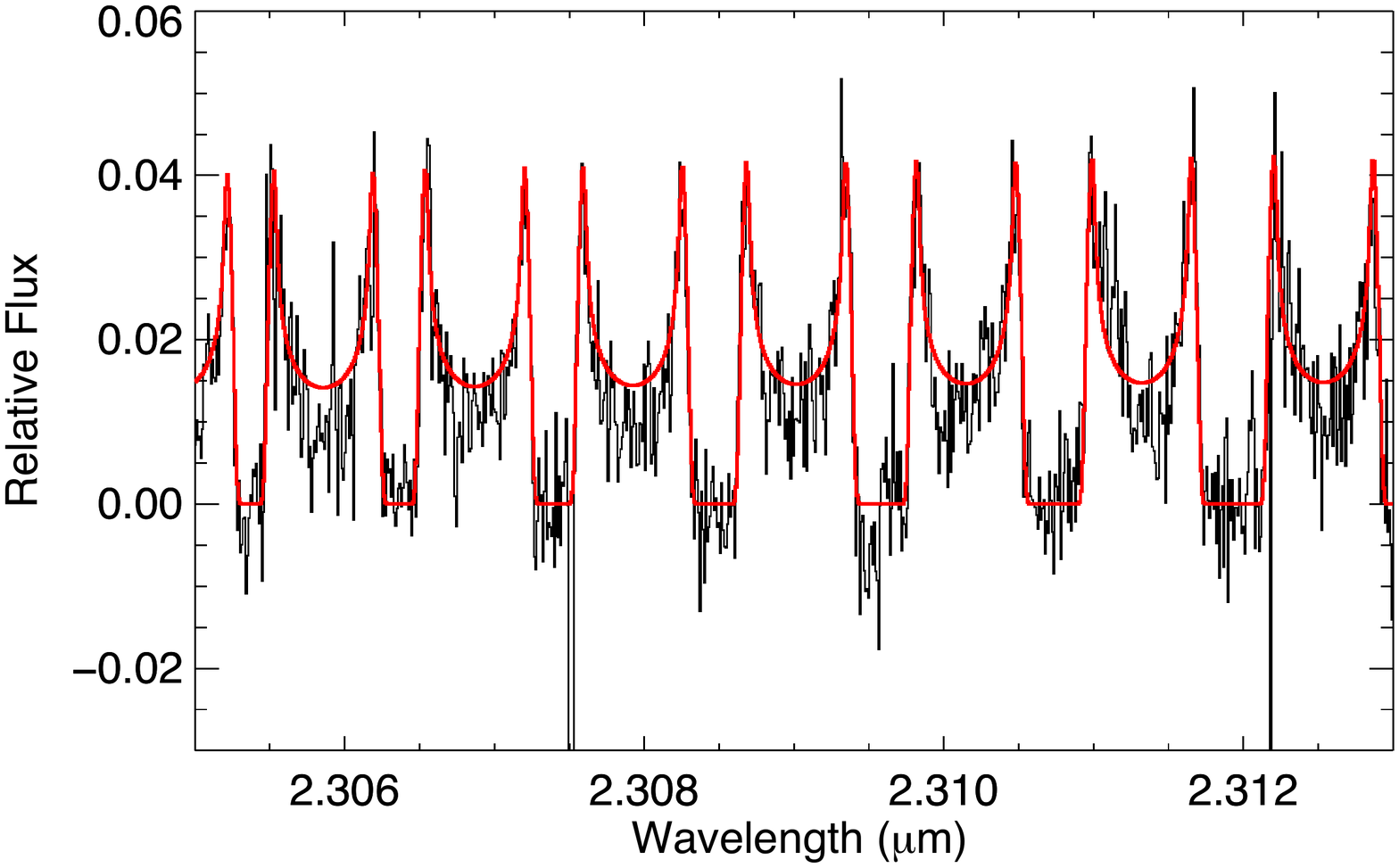}
\caption{
Isolated CO $\nu$ = 2 $\rightarrow$ 0 emission lines. The best fit model is plotted in red.
Final parameters for the model are T = 1300 K, N(CO) = 7 $\times$ 10$^{20}$ cm$^{-2}$,
with R$_{\rm{in}}$ = 0.88 and R$_{\rm{out}}$ = 1.27 AU, and i = 86$\degr$.
}
\label{fig:CO20iso}
\end{figure}

We start by fitting the velocity profile of the CO $\nu$ = 2 $\rightarrow$ 0 lines, because the CO spectrum covers the greatest range of energy levels and has the highest 
signal-to-noise. The CO lines near 2.31 $\mu$m (Figure \ref{fig:CO20iso}) are separated from other lines and give a clean measure of the line profile. A composite profile 
is formed using the four lines least affected by telluric absorption. A Keplerian disk emission model is fit to the profile using $\chi^{2}$ minimization, which gives 50.5 km 
s$^{-1}$ for the projected velocity at the inner radius of the emitting region, and 42.1 km s$^{-1}$ at the outer radius (or equivalently, R$_{\rm{out}}$/R$_{\rm{in}}$ = 1.44). 
A third fit parameter is the exponent of a power law for the radial intensity, I $\propto$ r$^{\alpha}$; however, the result is insensitive to this parameter, due to the small radial 
extent of the emission, and the exponent is set to a fixed value of $\alpha$ = $-$2.

\begin{figure*}
\includegraphics[width=\textwidth,trim={0.0in 1.8in 0.0in 2.in},clip]{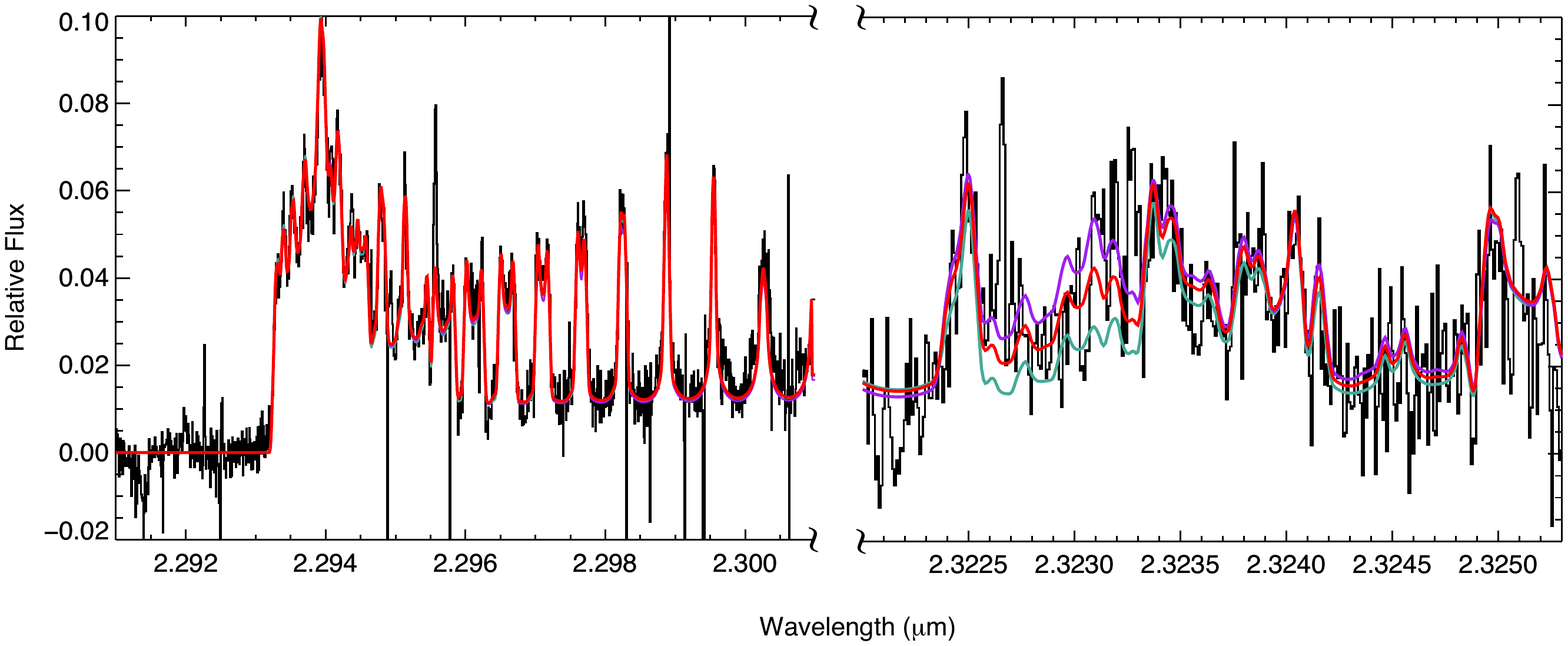}
\caption{
CO $\nu$ = 2 $\rightarrow$ 0 (left) and 3 $\rightarrow$ 1 (right) bandhead emission modeling. The best fit model is plotted in red. Final parameters for the red model are 
the same as in Figure \ref{fig:CO20iso}. Data is plotted in black, with the turquoise model indicating a temperature of 1000 K, red has a 
temperature of 1300 K, and purple has a temperature of 1500 K. The model is most sensitive to the region near 2.323$\mu$m. 
}
\label{fig:CObandmod}
\end{figure*}

Having set the kinematics of the CO emission, the CO $\nu$ = 2 $\rightarrow$ 0 bandhead and isolated lines are fit with a disk emission model. Given the narrow radial extent 
of the emission, the CO column density, N(CO), and temperature, T, are taken to be constant with radius. The main parameters that determine the bandhead shape and relative 
line intensities are T and N(CO). If you hold T constant, increasing N(CO) will increase the luminosity of the emission lines. For a given T and N(CO), matching the 
CO emission flux gives the {\it projected} emitting area, which is characterized by the radius of an equivalent 
circular area. The shape of the bandhead is also affected to a lesser degree by the local line broadening; the local line width was initially set to the CO thermal width. We find that 
the model fits to the CO $\nu$ = 2 $\rightarrow$ 0 bandhead have a large degeneracy between T and N(CO). Acceptable fit temperatures range from 1000 K to 2500 K, and we rule 
out emission for temperatures above 3000 K or below 800 K. When the CO $\nu$ = 3 $\rightarrow$ 1 bandhead is included in the fit, the relative flux of the $\nu$ = 3 $\rightarrow$ 1 
and $\nu$ = 2 $\rightarrow$ 0 emission restricts the range in T and N(CO), removing the degeneracy from fitting the CO $\nu$ = 2 $\rightarrow$ 0 alone  (Figure \ref{fig:CObandmod}).

The best fit parameters for the CO overtone emission are T = 1300$^{+200}_{-300}$ K and N(CO) = 7.0$^{+6.9}_{-1.3}$ $\times$ 10$^{20}$ cm$^{-2}$, and the projected emitting area, 
$\pi$R$_{\rm{e}}^{2}$, has a radius of R$_{\rm{e}}$ = 0.156 AU. This model is overplotted in red on the CO $\nu$ = 2 $\rightarrow$ 0 and $\nu$ = 3 $\rightarrow$ 1 bandheads in 
Figure \ref{fig:CObandmod}, along with models for higher and lower temperatures. 

Once we have the emitting column and projected area, we can break the degeneracy between disk inclination and radius by finding the inclination that is consistent with both 
the projected velocity and the projected emitting area. For the above solution, this inclination is i = 86$\degr$. The inner and outer radii for the CO emission are then are 
R$_{\rm{in}}$ = 0.88, R$_{\rm{out}}$ = 1.27 AU

\begin{figure}
\includegraphics[width=3.25in,height=2.25in,trim={0.25in 1.0in 1.0in 1.25in},clip]{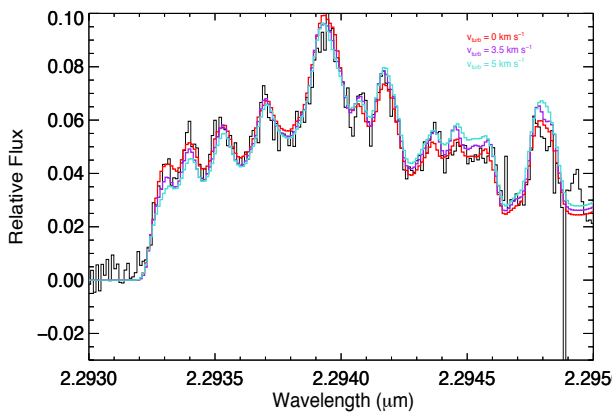}
\caption{
Models of the CO $\nu$ = 2 $\rightarrow$ 0 bandhead for different levels of turbulent broadening compared to the observed spectrum. The nominal model, with only thermal broadening, is 
plotted in red. The turquoise model includes an additional broadening (FWHM) of 5.0 km s$^{-1}$, which corresponds to the total linewidth of 5.2 km s$^{-1}$ in \cite{Ilee14}. We rule out 
v$_{\rm{turb}}$ (FWHM) $>$ 3.5 km s$^{-1}$ (purple plot).
}
\label{fig:turb}
\end{figure}

We also investigate the impact of non-thermal broadening (turbulence). The thermal width of CO at 1300 K is 1.5 km s$^{-1}$ (FWHM). Different amounts of extra broadening are added 
to the thermal width and the CO composite profile is refit. Then, the fits to the $\nu$ = 2 $\rightarrow$ 0 bandhead are repeated. A change in the local linewidth alters the overlap of the closely 
spaced transitions at the bandhead. Because the CO lines are optically thick, the amount of overlap affects the relative distribution of flux with wavelength and hence the shape of the bandhead. 
As the turbulence becomes larger, the fit to the shape at the bandhead  becomes progressively worse. Based on this, we rule out v$_{\rm{turb}}$ (FWHM) $\geq$ 3.5 km s$^{-1}$ (Figure 
\ref{fig:turb}).

In modeling the OH, we first determine the radial extent of the OH emission by modeling the profile of the blended OH P4.5 doublet feature, using the same procedure used for the CO profile. 
Using the same inclination angle (86$\degr$) from the CO modeling, the OH emitting region extends from R$_{\rm{in}}$ = 0.81 to R$_{\rm{out}}$ = 1.46 AU. Figure \ref{fig:ohall}, panel A shows 
the model fit to the P4.5 doublet. The model includes non-thermal line broadening (FWHM) of 6.7 km s$^{-1}$, which improves the appearance of the fit at the peaks of the OH emission; however, 
the statistical significance, vs. thermal broadening, is small, and its inclusion does not change the derived radii for the emission. The same best fit velocity profile is consistent with the P5.5 
emission feature, as show in Figure \ref{fig:ohall}, panel C.

The OH and CO emission originate from similar radii, but the radial extent (and area) of the OH emission is somewhat larger than that found for the CO emission. Given that the CO and OH 
spectra were obtained 2 years apart, it is not clear whether the OH and CO line profiles point due to an intrinsic difference in their respective radial distributions or reflect the 
variability of the emitting size.  

In order to derive a column density for the OH emission, we adopt the temperature of 1300 K found for CO, since the the OH features give no constraint on the gas temperature. Using the 
projected area for the OH emission, the column density is adjusted to match the flux in the OH P4.5 doublet. We find that N(OH) = 2.8$^{+1.4}_{-0.7}$ $\times$ 10$^{18}$ cm$^{-2}$, which 
yields a ratio of N(OH)/N(CO) = 4.0$^{+2.3}_{-2.5}$ $\times$ 10$^{-3}$.

\begin{figure*}
\includegraphics[width=\textwidth,trim={1.0in 2.in 1.2in 2.4in},clip]{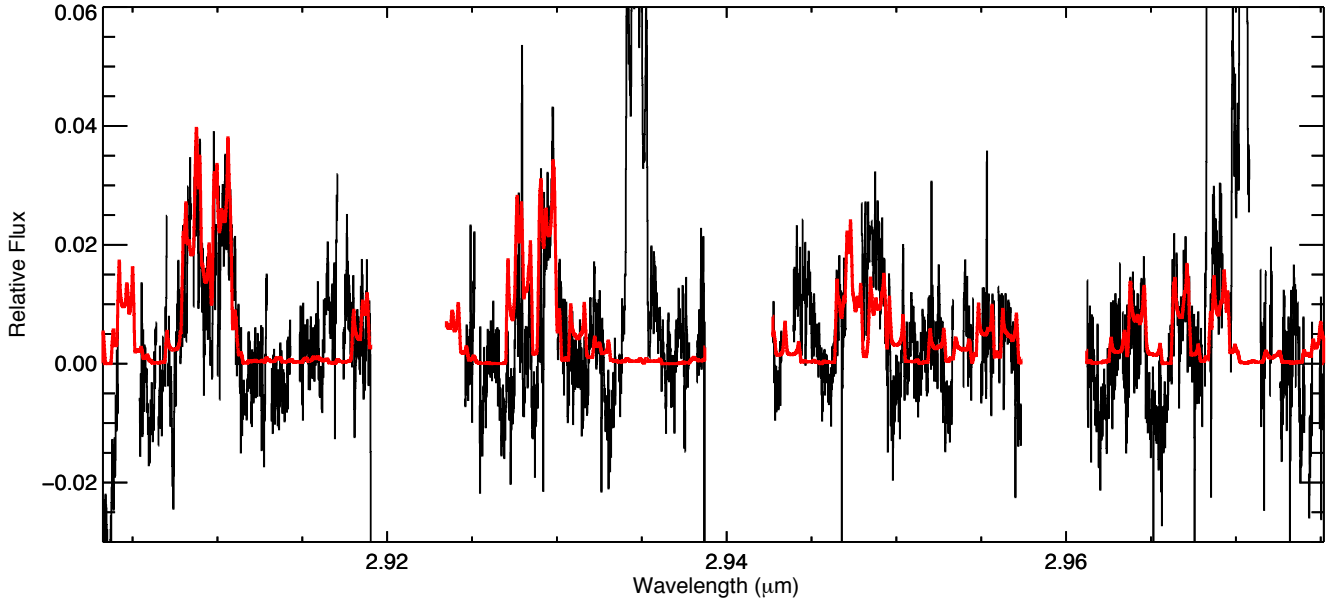}
\caption{
Model of observed H$_{2}$O emission features. The model uses the same geometry as the best fit OH P4.5 model: R$_{\rm{in}}$ = 0.81 AU, R$_{\rm{out}}$ = 1.46 AU, and an inclination 
angle of 86$\degr$.
}
\label{fig:JohnH2O}
\end{figure*}

Modeling of the H$_{2}$O emission is more complicated. We originally confirmed our identification of these features as water by comparison to emission from LTE slab models. Due to the 
lower signal-to-noise of the H$_{2}$O emission, we find that it is not possible to uniquely determine the temperature and column density of water from the spectrum, although it is clearly hot, 
in the range of 1000 to 3000 K. In addition, the velocity line profile can not be constrained to the accuracy that is possible for CO and OH. Hence, the OH velocity profile and emitting area are 
used for H$_{2}$O, along with the same 1300 K temperature. The column density required to match the H$_{2}$O flux is N(H$_{2}$O) = 5.8$^{+0.6}_{-0.8}$ $\times$ 10$^{17}$ cm$^{-2}$. 
This model is compared to the H$_{2}$O emission features in Figure \ref{fig:JohnH2O}. Other features, outside of those mentioned in Section 3.2, are consistent with the H$_{2}$O emission 
model. The adopted parameters are consistent with the relative fluxes and velocity widths in the H$_{2}$O spectrum. The derived water column density yields ratios of N(H$_{2}$O)/N(OH) 
= 0.21$^{+0.11}_{-0.06}$ and N(H$_{2}$O)/N(CO) = 8.3$^{+3.0}_{-4.1}$ $\times$ 10$^{-4}$.

\section{Comparison to other Young Stellar Objects}

To contextualize our detection of water in the inner disk around HD~101412, we compare the column densities of CO, OH, and H$_2$O among seven other young stars for which water has been detected (HD~259431 does not have a detection of NIR water emission, just an upper limit on the water column density) (Table \ref{tab:Comp}). SVS~13 is a $\geq3$M$_{\sun}$ \citep{Hirota08} young stellar object on the Class 0/Class I boundary \citep{Chen09} from which the CO $\nu$ = 2 $\rightarrow$ 0 bandhead and H$_{2}$O emission lines near 2.2935 $\mu$m have been observed \citep{Carr04}. AS~205A, DR~Tau, and RU~Lup are all classical T Taruri stars (CTTS; spectral type K0, K5, and G5, respectively) around which CO, OH, and H$_{2}$O emission has been observed \citep{Salyk08,Mandell12}. V1331~Cyg is a 2.8~M$_{\sun}$ intermediate mass T Tauri star (IMTTS; spectral type G7-K0; \citealt{Petrov14}). \citet{Doppmann11} observed OH and H$_{2}$O emission in the $L$-band from this star. 08576nr292 is a massive young stellar object (MYSO; $\sim$6 M$_{\odot}$ B5 star) from which H$_{2}$O and CO bandhead emission have also been detected \citep{Thi05}. 

For four of the sources (HD~101412, SVS~13, V1331~Cyg, and 08576nr292) in Table \ref{tab:Comp}, CO column density is comparable (0.7-6 $\times$ 10$^{21}$ molecules cm$^{-1}$). Three other source (AS~205A, DR~Tau, and HD~259431) have much lower CO column density reported, ranging from 0.6-1.6 $\times$ 10$^{19}$ molecules cm$^{-1}$ (RU~Lup only has column density ratios reported in \citealt{Mandell12}). However, even with the range in column densities, the ratios between molecules show similar trends when comparing similar sources. Lower mass T Tauri stars have N(OH)/N(CO) of a 2-3.3 $\times$ 10$^{-2}$ and N(H$_{2}$O)/N(CO) of 0.6-3.0 $\times$ 10$^{-1}$. As you move to more massive sources, this trend changes. N(OH)/N(CO) is now 0.5-4 $\times$ 10$^{-3}$ and N(H$_{2}$O)/N(CO) is 0.2-8 $\times$ 10$^{-4}$. The fact that the latest type star is surrounded by the disk with the highest H$_{2}$O/CO ratio suggests 
that the UV radiation from the star plays a pivotal role in determining the abundance of water in the atmosphere of the inner disk 

In addition to the comparison among a handful of young stars for which water has been detected, we compare a sample of 19 T Tauri and 
HAeBe stars reported in the literature for which both OH and H$_{2}$O are measured (Table 4; \citealt{Fedele11,Banzatti17}). Plotting the OH line luminosity 
versus the H$_2$O line luminosity, we find that the T Tauri stars follow a linear trend given by Eqn. \ref{eqn:LumTrend}, confirming the constant OH/H$_{2}$O line 
flux ratios presented in \cite{Banzatti17},

\begin{equation} \label{eqn:LumTrend}
L_{H_2O}/L_{\sun}=(2.1\pm0.1)L_{OH}/L_{\sun} + (0.09\pm0.23).
\end{equation}

\begin{deluxetable*}{lcccccccc}[t]
\tabletypesize{\scriptsize}
\tablecaption{Column Density Comparison to other YSOs}
\tablewidth{0pt}
\tablehead{
\colhead{Star} & \colhead{SpT} & \colhead{M$_{\star}$} & \colhead{Class} & \colhead{N(CO)} & \colhead{N(OH)} & \colhead{N(H$_{2}$O)} & \colhead{N(OH)/N(CO)} & \colhead{N(H$_{2}$O)/N(CO)}\\
\colhead{} & \colhead{} & \colhead{(M$_{\sun}$)} & \colhead{} & \colhead{(cm$^{-2}$)} & \colhead{(cm$^{-2}$)} & \colhead{(cm$^{-2}$)} & \colhead{} & \colhead{}
}
\startdata
SVS~13$^{1}$ & -- & 3.0 & Class 0/1 & 1.2 $\times$ 10$^{21}$ & -- & 1.8 $\times$ 10$^{21}$  & -- & 1.5\\
DR Tau$^{2}$ & K5 & 0.8 & CTTS & 7.0 $\times$ 10$^{18}$& 2.0 $\times$ 10$^{17}$ & 8.0 $\times$ 10$^{17}$ & 2.9 $\times$ 10$^{-2}$ & 1.1 $\times$ 10$^{-1}$\\
AS 205A$^{2}$ & K0 & 1.2 & CTTS & 6.0 $\times$ 10$^{18}$ & 2.0 $\times$ 10$^{17}$ & 6.0 $\times$ 10$^{17}$ & 3.3 $\times$ 10$^{-2}$ & 1.0 $\times$ 10$^{-1}$\\
V1331~Cyg$^{3}$ & G7-K0 & 2.8 & IMTTS & 6.0 $\times$ 10$^{21}$ & 1.0 $\times$ 10$^{20}$ & 2.0 $\times$ 10$^{21}$ & 2.0 $\times$ 10$^{-2}$ & 3.0 $\times$ 10$^{-1}$\\
RU Lup$^{4}$ & G5 & 0.7 & CTTS & -- & -- & -- & 1.6 - 3.3 $\times$ 10$^{-2}$ & 5.5 $\times$ 10$^{-2}$ \\
HD~101412 & B9.5 & 2.5 & HBe & 7.0$^{+6.9}_{-1.3}$ $\times$ 10$^{20}$ & 2.8$^{+1.4}_{-0.7}$ $\times$ 10$^{18}$ & 5.8$^{+0.6}_{-0.8}$ $\times$ 10$^{17}$ & 4.0$^{+2.3}_{-2.5}$ $\times$ 10$^{-3}$ & 8.3$^{+3.0}_{-4.1}$ $\times$ 10$^{-4}$\\
08576nr292$^{5}$ & B5 & 6.0 & MYSO & 3.9 $\times$ 10$^{21}$ & -- & 2.5 $\times$ 10$^{18}$ & $\sim$1.0 $\times$ 10$^{-3}$ & 6.4 $\times$ 10$^{-4}$ \\
HD 259431$^{6,7}$ & B5 & 6.6 & HBe & 1.6 $\times$ 10$^{19}$ & 7.9 $\times$ 10$^{15}$ & $<$3.2 $\times$ 10$^{14}$ & 4.9 $\times$ 10$^{-4}$ & $<$2.0 $\times$ 10$^{-5}$
\enddata
\label{tab:Comp}
\tablecomments{
Comparison of molecular column densities from previously reported observations of young stellar objects. \citet{Thi05} did not observe OH emission in 08576nr292. N(OH)/N(CO) 
for 08576nr292 is based off chemical models. 1. \cite{Carr04}; 2. \cite{Salyk08}; 3. \cite{Doppmann11}; 4. \cite{Mandell12}; 5. \cite{Thi05}; 6. \cite{Ilee14}; 7. \cite{Fedele11};.
}
\end{deluxetable*}

\noindent Among the sources included in this sample is EX Lupi while it was undergoing an outburst and in quiescence \citep{Banzatti17}. 
The lumiosity increases along the fit to the T Tauri data indicating that the ratio of L$_{\rm{H_{2}O}}$ and L$_{\rm{OH}}$ is relatively constant over a wide 
range of stellar luminosities. However, it is not clear if the FUV luminosity of the star would impact the emission from the circumstellar disk while undergoing an 
outburst. During outbursts, the inner disk heats up to the point that the continuum emission from the inner region buries the emission from the star. The outer disk would 
thus only see emission from the self-luminous inner disk. For example, the FUV spectra of T Tauri stars is more similar to FUors than to the far more FUV luminous HAeBes 
\citep{Valenti00}.

While HD~101412 is the only HAeBe in the sample for which both water and OH are detected, upper limits for eight additional sources are available from 
\cite{Fedele11}. We find that the HAeBes consistently show weaker water luminosity for a given OH luminosity than the T Tauris. This trend is also suggestive 
that the UV luminosity of the stars plays an important role in determining the relative column density of water. 
 
\begin{figure}
\includegraphics[width=3.3in,height=2.9in,trim={2.in 1.in 2.5in 1.5in},clip]{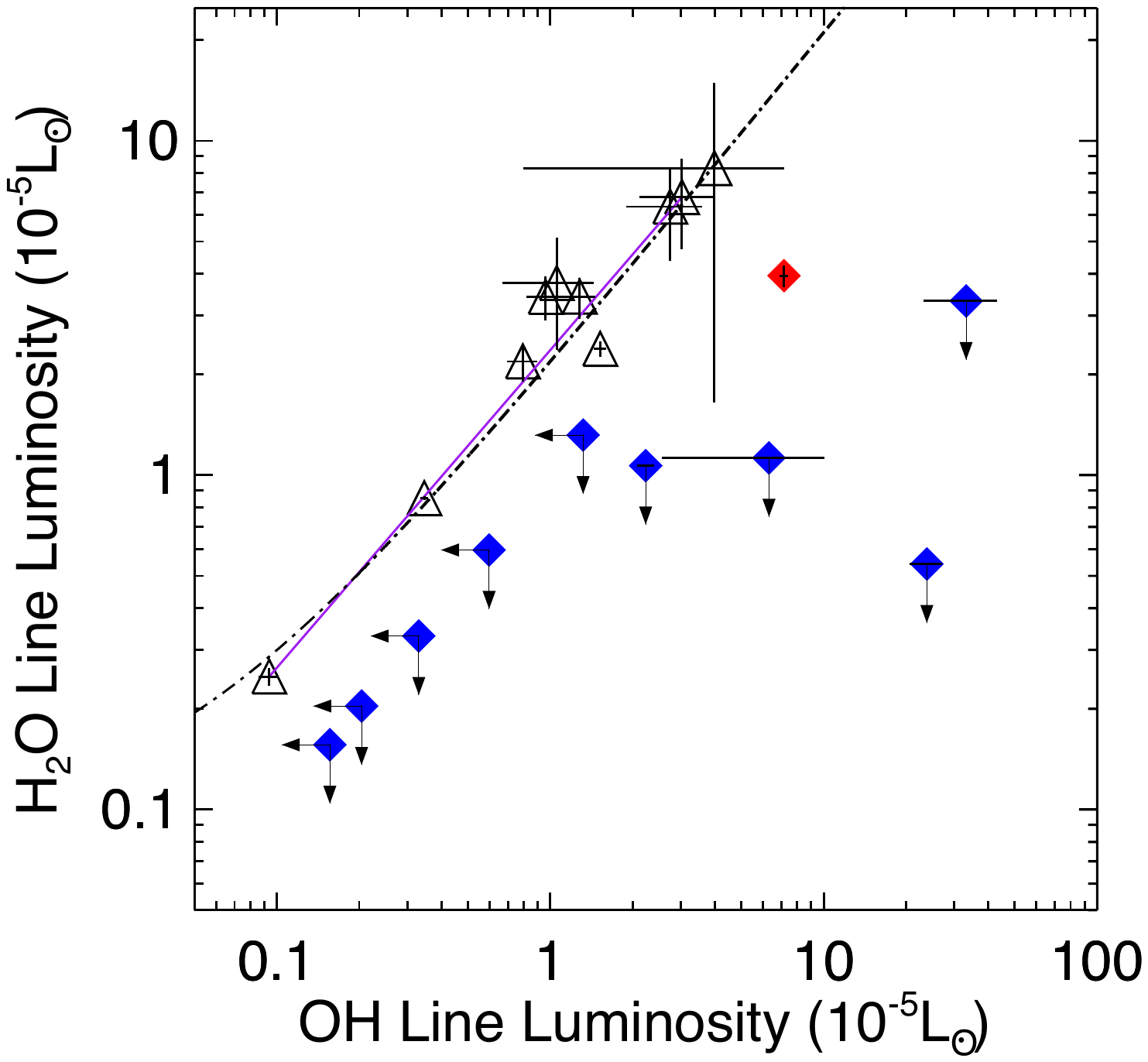}
\caption{
Comparison of the H$_{2}$O line luminosity to the OH line luminosity taken from literature \citep{Banzatti17,Fedele11} for T Tauri and HAeBe stars. T Tauri 
stars are plotted as open black triangles, HAeBe stars as blue diamonds, and HD~101412 as a red diamond. The black line is a weighted linear least squares fit to the T Tauri 
data. We do not plot T Tauri sources with upper limits from \cite{Banzatti17}, only T Tauris with detections of both water and OH at 2.9 $\mu$m. We only plot the HAe/Be 
sources with OH detections or upper limits and water upper limits for 2.9 $\mu$m from \cite{Fedele11}, and we have converted HAeBe upper limits to 1$\sigma$ levels for consistency.
The purple line connects EX Lup observations from 2008 and 2014. The 2008 observation occurred during an accretion outburst resulting in higher OH and H$_{2}$O fluxes. The luminosity 
of the water emission from the T Tauri stars is consistently larger than the upper limits placed on HAeBe stars.
}
\label{fig:OHH2O}
\end{figure}

\section{Discussion}

\begin{deluxetable*}{lccccc}[t]
\tabletypesize{\scriptsize}
\tablecaption{H$_{2}$O Line Luminosity vs. OH Line Luminosity for YSOs}
\tablewidth{0pt}
\tablehead{
\colhead{Star} & \colhead{SpT} & \colhead{M$_{\star}$} & \colhead{Class} & \colhead{L$_{\rm{H_{2}O}}$} & \colhead{L$_{\rm{OH}}$} \\
\colhead{} & \colhead{} & \colhead{(M$_{\sun}$)} & \colhead{} & \colhead{(10$^{-5}$ L$_{\sun}$)} & \colhead{(10$^{-5}$ L$_{\sun}$)}
}
\startdata
AS 205A 		&	 K0 		&	1.2	&	 CTTS 	&	3.76	$\pm$	1.39	&	1.06	$\pm$	0.39		\\
DF Tau 	    	&  	 K5 		&	0.8	&	 CTTS 	&	2.19	$\pm$	0.28	&	0.79	$\pm$	0.10		\\
DR Tau 	    	&	 M3 		&	0.5	&	 CTTS 	&	3.41	$\pm$	0.52	&	0.96	$\pm$	0.14		\\
EX Lup08 		&	 M0 		&	0.8	&	 CTTS 	&	6.79	$\pm$	2.04	&	3.03	$\pm$	0.91		\\
EX Lup14 		&	 -- 		&	 -- 	&	 -- 		&	0.25	$\pm$	0.02	&	0.09	$\pm$	0.01		\\
RU Lup  		&	 K7-M0 	&	0.7	&	 CTTS 	&	3.42	$\pm$	0.48	&	1.28	$\pm$	0.18		\\
S Cra N 		&	 G0 		&	0.6	&	 CTTS 	&	6.36	$\pm$	1.97	&	2.75	$\pm$	0.85		\\
T Tau N 		&	 K1.5 	&	2.4	&	 IMTTS 	&	8.28	$\pm$	6.63	&	3.98	$\pm$	3.18		\\
VW Cha  		&	 K7 		&	0.6	&	 CTTS 	&	2.39	$\pm$	0.13	&	1.52	$\pm$	0.08		\\
VZ Cha  		&	 K7 		&	0.8	&	 CTTS 	&	0.85	$\pm$	0.01	&	0.35	$\pm$	0.01		\\
BF Ori  		&	 A5 		&	1.4	&	 HAe 	&	$<$0.16		        	&	$<$0.16				\\
HD 34282  	&	 A0 		&	1.9	&	 HAe 	&	$<$0.20		        	&	$<$0.20				\\
HD 76534 	&	 B2 		&	11.4  &	 HBe 	&	$<$1.51		        	&	$<$1.51				\\
HD 85567 	&	 B5 		&	6	&	 HBe 	&	$<$3.32			&	33.19	$\pm$	9.96  \\
HD 98922 	&	 B9 		&	5.2	&	 HBe 	&	$<$0.40		        	&	$<$0.40				 \\
HD 101412 	&	 B9.5 	&	2.5	&	 HBe 	&	3.96	$\pm$	0.30 &	7.18	$\pm$	0.25		 \\
HD 250550 	&	 B7 		&	3.6	&	 HBe 	&	$<$1.06			&	2.23	$\pm$	0.16		 \\
HD 259431 	&	 B5 		&	6.6	&	 HBe 	&	$<$0.55		        	&	23.86	$\pm$	3.29	 \\
UX Ori  		&	 A3 		&	2.1	&	 HAe 	&	$<$0.33		        	&	$<$0.33				 \\
V380 Ori 		&	 A1 		&	2.8	&	 HAe 	&	$<$1.13		        	&	6.31	$\pm$	3.75	
\enddata
\label{tab:FluxComp}
\tablecomments{
Luminosity values from literature used in Figure \ref{fig:OHH2O}. T Tauri flux values are obtained from \cite{Banzatti17}, and HAeBe flux values and upper limits from \cite{Fedele11}. 
Flux values are converted to luminosities using distances obtained from \cite{Gaia16}. Upper limits have been converted to 1$\sigma$ limits for consistency.
}
\end{deluxetable*}

\begin{figure*}[t]
\plotone{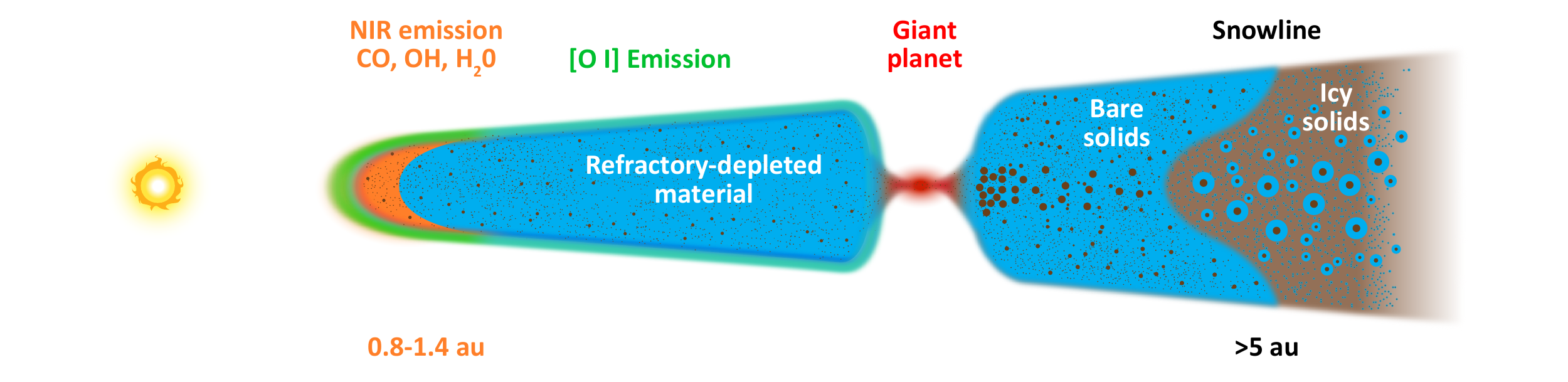}
\caption{
The above schematic presents one possible scenario, discussed in detail in the text, to explain the NIR observations of HD~101412.
Based on the stellar luminosity for HD~101412, we determine the dust sublimation radius to be at 0.25 AU. The molecular emission arises 
from a narrow annulus ($\sim$0.8 - 1.4 AU). Refractory elements are observed to be below solar abundances based on photospheric observations, while C and 
O are found to be at solar abundances \citep{Folsom12}. This indicates that large dust grains are filtered out, leading to a depletion of refractory 
elements in the inner disk. If a planetary body was forming within the snow line ($>$5 AU), this would allow for C- and O-containing molecules to reside 
in the inner disk, eventually accreting onto the central star and giving rise to the observed photospheric abundance pattern. This figure is not to scale.
}
\label{fig:DiskScheme}
\end{figure*}

\begin{figure}
\includegraphics[width=3.35in,height=2.15in,trim={0.5in 1.25in 0.75in 1.4in},clip]{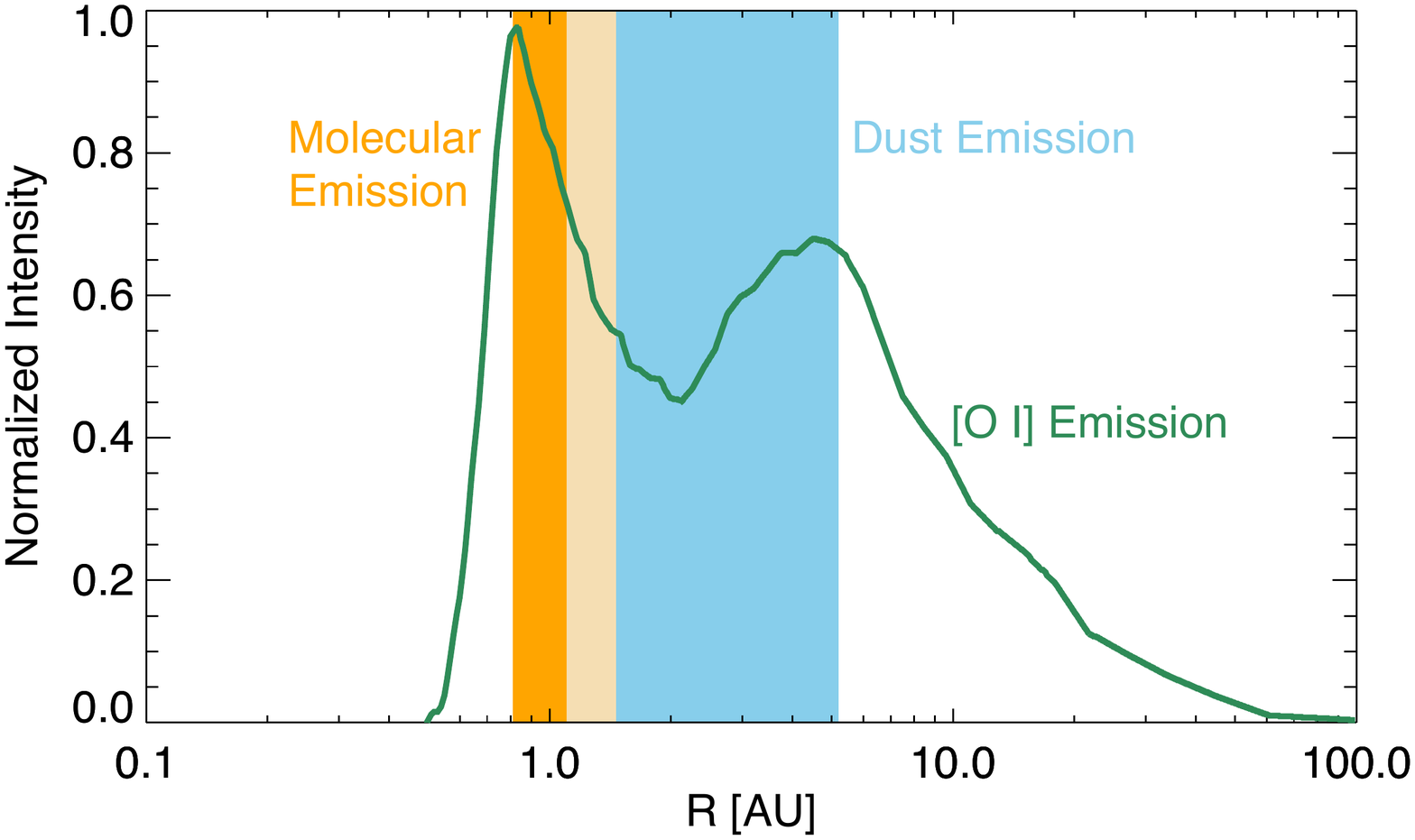}
\caption{
Updated Figure 8 from \cite{Fedele08} showing locations of [\ion{O}{1}] emission, dust emission, and molecular emission from HD~101412. The orange indicates the 
location of OH, H$_{2}$O, and CO gas presented from our analysis. The blue region indicates the dust emission based on MIDI observations, updated to account for the 
\citet{Gaia16} distance. The light brown is the overlap of the two regions, with the [\ion{O}{1}] emission (in green) remaining unchanged due to the use of similar 
stellar masses between the studies.
}
\label{fig:MIDI}
\end{figure}

The infrared molecular emission from HD~101412 is unusual in several respects. Firstly, we see 
the CO bandhead emission arising from a narrow annulus. To populate the CO bandheads, the gas must be hot (T $\gtrsim$ 2000~K) and dense 
(n$_H\gtrsim10^{10}$~cm$^{2}$; \citealt{Najita96}). The requisite conditions are ordinarily 
only met in systems with high accretion rates ($\sim$10$^{-7} - 10^{-6}$~M$_{\odot}$yr$^{-1}$, \citealt{Ilee14}).  
CO bandhead emission is rarely observed in HAeBe systems, with a detection rate of 7\% \citep{Ilee14}.

In order to detect the large columns of CO gas observed in emission, 
all CO bandhead sources require strong suppression of the $K-$band opacity 
in the CO-emitting region.  For HD~101412, 
the CO column density inferred from overtone bandhead emission (7$\times$10$^{20}$~cm$^{-2}$; Table 3) corresponds to 
N$\rm_H$=1.4$\times$10$^{25}$~cm$^{-2}$, assuming a CO/H$_{2}$ of 1 $\times$ 10$^{-4}$ or an A$_{\rm{K}}$ = 600 
if the dust were interstellar.  Detecting a CO column as large as $7\times 10^{20}$~cm$^{-2}$ therefore 
requires a reduction in the K-band continuum opacity by a factor of 
$\sim$600, i.e., a factor of $\sim$6 larger than the typical 
factor of $\sim$100 reduction 
in grain surface area that is found for T Tauri disks 
\citep{Furlan07}.

\cite{Ilee14} mention that CO emission is primarily observed around B-type stars, which makes sense due to the required temperatures 
to excite CO bandhead emission. Conditions around lower mass stars may only reach requisite temperatures during episodic accretion 
events thus resulting in variable CO bandhead emission. Also, due to the high gas density required, some disks may lack sufficient 
material to allow for CO bandhead emission. \cite{Ilee14} also mention a possible connection to high disk inclinations with CO bandhead 
emission. Their sources with detections had a range of inclinations from 51$\degr$ to 72$\degr$, based on model fits. A possible 
explanation for this inclination dependance would be the CO bandhead emission tracing the inner disk wall near the dust sublimation radius.

Secondly, whereas water emission is rare among HAeBe disks (Section 1), 
we detect in HD~101412 hot water emission with a luminosity 
comparable to the most luminous water emission from T Tauri disks. 
One possible explanation for the dearth of 
water emission from HAeBe disks compared to T Tauri disks is the relative UV luminosities to which the circumstellar 
disks are exposed: the strong FUV field of HAeBe stars can readily dissociate water in their disk atmospheres. This may not be apparent from 
the \citet{Walsh15} chemical model of a disk around a HAe star, which 
finds a water-rich disk atmosphere with 
water column densities much larger than is consistent with observations. As they note, one reason for the discrepancy between the observed and predicted 
water columns 
may be their assumption of interstellar grains (gas-to-dust ratio and grain size distribution), which limits the penetration depth 
of the UV photons and their effect on disk molecular abundances.

The T Tauri disk models of \citet{Adamkovics14} (their Figure 4; see also \citet{Adamkovics16} and \citet{Najita17}) 
support this perspective. Assuming grain growth at 
the level inferred for observed sources (e.g., \citealt{Furlan07}), these models find that increasing the 
FUV radiation from T Tauri 
stars does in fact push the molecular layer deeper into the disk and dissociates H$_2$O to produce more extensive 
OH. Models of HAeBe disks that assume a comparable level of grain growth, 
would likely find a similar reduction in the water column density in the disk atmosphere, more  
consistent with the general lack of water emission detected from HAeBe stars (Figure \ref{fig:OHH2O}). In principle, 
water emission could be detected in HAeBe disks if the dust opacity in the disk atmosphere was low enough that the 
dust photosphere was located below the transition from OH to H$_2$O.

There are reasons to expect a low dust opacity in the inner disk of HD~101412. The stellar photosphere of HD~101412 
is strongly depleted in refractory elements (Fe, Mg, Si), but has solar-like abundances of volatile elements (C, N, O), 
thus, HD~101412 is a $\lambda$ Bo\"o star \citep{Folsom12}. \citet{Kama15} hypothesize that the depletion of heavy 
elements in the photosphere of the star is a consequence of selective accretion of gas relative to dust, that is the 
accreting material accreting has a gas-to-dust ratio of $\sim$600, i.e., a reduction in refractories by a factor of $\sim 6$, similar 
to the extra factor of 6 reduction needed in the K-band continuum opacity 
to expose the entire CO bandhead-emitting column to view.

What is the source of the depletion of refractory material in the disk? If a giant planet (with mass from 0.1 M$_{\rm{J}}$ 
to 10 M$_{\rm{J}}$) is present in the disk, the pressure bumps it creates (e.g., at the edge of a gap) could preferentially 
reduce the accretion of solids relative to gas through the disk by aerodynamic drag \citep{Rice06,Zhu12}. Due to the lack 
of a convective outer layer in early-type stars, the refractory depleted material that accretes onto the star would reside on 
the surface, giving rise to the $\lambda$ Bo\"o abundance pattern (Figure \ref{fig:DiskScheme}).

The detection of C- and O-bearing molecules (CO, OH, and H$_{2}$O) from the inner region of the HD101412 
disk and the detection of CO$_2$ with Spitzer \citep{Pontoppidan10}, which presumably will eventually accrete onto the star, is consistent 
with the high abundance of volatile elements in the stellar atmosphere of HD~101412. The presence of oxygen-bearing  
molecules such as H$_{2}$O and CO$_{2}$ in the gas phase implies that if a planet induced gap is responsible for filtering 
out the solids from the inwardly accreting material \citep{Kama15}, the planet would likely be located well inward of the snow 
line (which we estimate is located at $\sim$25\,AU adopting the distance at which a blackbody is 150K, but must lie beyond 5 au; 
Figure \ref{fig:DiskScheme}).  If the planet were located near or beyond the snow line, dust filtering would remove water ice 
(and oxygen) as well from the accreting material, creating a more carbon-rich composition, in which molecules like H$_{2}$O 
and CO$_{2}$ are unlikely to be abundant (Figure \ref{fig:DiskScheme}). 

The NIR molecular emission from 
HD~101412 overlaps the [\ion{O}{1}] emission from the source and is roughly coincident spatially with the inner edge 
of the dust disk seen in the mid-infrared. The mid-infrared dust emission arises from 1.1 - 5.2 AU when taking 
into consideration the new distance from Gaia (Figure \ref{fig:MIDI}; \citealt{Fedele08,vanderPlas08}).The structure 
shown in Figure \ref{fig:MIDI} is analogous to a miniature photo dissociation region with the inner emission arising from [\ion{O}{1}], then molecular emission, 
followed by dust emission. The [\ion{O}{1}] emission probes the tenuous atomic layer of the disk that is depleted of dust 
and molecules. It is not clear what the radial extent of the disk is as there are presently no far-infrared or submillimeter 
observations that would probe the cooler dust arising from beyond 5 AU. 

Given the high inclination of the system (i = 86$\degr$; Section 4), the hot molecular emission we see may be coming from the 
inner wall of the far side of the disk. A viewing angle 4$\degr$ from edge-on appears large enough 
to obtain an unobstructed view of the bright molecular emission from the far side of the inner disk. Given the temperature of 
the molecular emission (1300 K; Section 4) and the stellar mass (2.5 M$_{\odot}$; Section 1), the disk  scale height at the radius 
of the molecular emission (1 AU) is 0.05 AU if the gas is in hydrostatic equilibrium; thus the height of the molecular emitting 
gas on the near side of the disk may rise $\sim$1.5$\degr$ above the midplane when viewed from the far side of the disk 2~AU 
away. 

There are two observational approaches to test our hypothesis that the luminosity of the water emission and CO bandhead 
emission are enhanced by the depletion of dust in the inner disk (by a factor of 6 or more; \citealt{Kama15}). Firstly, one can compare
 the equivalent width of the molecular emission to the gas-to-dust ratio of the accreting material 
as inferred from photospheric abundances of the star. Larger molecular emission equivalent widths are expected for systems 
with larger gas-to-dust ratios. Secondly, one can test the role of the disk inclination by observing additional edge-on systems 
and comparing these to more inclined systems. The emission from the edge-on systems will be dominated by the inner wall 
of the disk and should probe denser gas than the more face-on systems where the emission is dominated by less dense gas 
from the disk surface.

The feasibility of this scenario can also be tested through thermochemical modeling of the inner disk. One possible concern 
is that if grains are too under-abundant, it is difficult to synthesize H$_2$ on grains, a critical first step in the gas phase 
synthesis of molecules such as OH, CO and water. Severe dust depletion may make it difficult to counteract photodestruction 
of molecules and sustain a substantial reservoir of molecular gas. Self-consistently fitting the infrared portion of the SED and 
molecular column density in the disk atmosphere will inform the feasibility of our hypothesis.

\section{Conclusions}

We present the first detection of water emission at 2.9 $\mu$m in a Herbig Ae/Be star system, along with a new detection of OH ro-vibrational emission. The OH emission observed 
represents the strongest ever observed in a HAeBe disk in terms of line-to-continuum ratio. The observed line profiles for both OH and H$_{2}$O indicate that the emitting region 
for both molecules is narrow and $\sim$1 AU from the star. 

The bright molecular emission from HD~101412 may be related to its photospheric abundance pattern, i.e., its nature as a $\lambda$ Bo\"o star, and the disk's large inclination angle. 
If the low abundance of refractory elements is a result of selective accretion of gas relative to dust, as has been previously hypothesized, the inner disk from which HD~101412 accretes 
should be strongly dust-depleted and its continuum more optically thin. This situation would tend to produce strong molecular emission from the inner disk, as is observed. Our detection 
of C- and O-bearing molecules from the inner disk is consistent with the expected presence in this scenario of abundant volatiles in the accreting material. 

\acknowledgements

The authors would like to thank the referee for comments which improved the content of the manuscript. Based also in part on data obtained from the ESO Science Archive Facility. 
S. D. B. and S. C. A. acknowledge support for this work from the National Science Foundation under grant number AST-1517014. M\'A is supported under 
NSF grant number AST-1517785. Basic research in infrared astronomy at the Naval Research Laboratory is supported by 6.1 base funding. S.D.B. also acknowledges 
support for this work from NASA Agreement Nos. NXX15AD94G and NNX16AJ81G and the stimulating environment provided by NOAO during his visit. NOAO is operated 
by the Association of Universities for Research in Astronomy (AURA), Inc. under a cooperative agreement with the National Science Foundation. 

{\it Facilities:} \facility{Very Large Telescope:UT1 (CRIRES)}

\bibliographystyle{apj}
\bibliography{101412cite}		

\begin{thebibliography}{}
\expandafter\ifx\csname natexlab\endcsname\relax\def\natexlab#1{#1}\fi

\bibitem[{{{\'A}d{\'a}mkovics} {et~al.}(2014){{\'A}d{\'a}mkovics}, {Glassgold},
  \& {Najita}}]{Adamkovics14}
{{\'A}d{\'a}mkovics}, M., {Glassgold}, A.~E., \& {Najita}, J.~R. 2014, \apj,
  786, 135

\bibitem[{{{\'A}d{\'a}mkovics} {et~al.}(2016){{\'A}d{\'a}mkovics}, {Najita}, \&
  {Glassgold}}]{Adamkovics16}
{{\'A}d{\'a}mkovics}, M., {Najita}, J.~R., \& {Glassgold}, A.~E. 2016, \apj,
  817, 82

\bibitem[{{Banzatti} \& {Pontoppidan}(2015)}]{Banzatti15}
{Banzatti}, A., \& {Pontoppidan}, K.~M. 2015, \apj, 809, 167

\bibitem[{{Banzatti} {et~al.}(2017){Banzatti}, {Pontoppidan}, {Salyk},
  {Herczeg}, {van Dishoeck}, \& {Blake}}]{Banzatti17}
{Banzatti}, A., {Pontoppidan}, K.~M., {Salyk}, C., {et~al.} 2017, \apj, 834,
  152

\bibitem[{{Blake} \& {Boogert}(2004)}]{Blake04}
{Blake}, G.~A., \& {Boogert}, A.~C.~A. 2004, \apjl, 606, L73

\bibitem[{{Brittain} {et~al.}(2016){Brittain}, {Najita}, {Carr},
  {{\'A}d{\'a}mkovics}, \& {Reynolds}}]{Brittain16}
{Brittain}, S.~D., {Najita}, J.~R., {Carr}, J.~S., {{\'A}d{\'a}mkovics}, M., \&
  {Reynolds}, N. 2016, \apj, 830, 112

\bibitem[{{Brittain} {et~al.}(2007){Brittain}, {Simon}, {Najita}, \&
  {Rettig}}]{Brittain07}
{Brittain}, S.~D., {Simon}, T., {Najita}, J.~R., \& {Rettig}, T.~W. 2007, \apj,
  659, 685

\bibitem[{{Brown} {et~al.}(2013){Brown}, {Pontoppidan}, {van Dishoeck},
  {Herczeg}, {Blake}, \& {Smette}}]{Brown13}
{Brown}, J.~M., {Pontoppidan}, K.~M., {van Dishoeck}, E.~F., {et~al.} 2013,
  \apj, 770, 94

\bibitem[{{Carr} {et~al.}(2004){Carr}, {Tokunaga}, \& {Najita}}]{Carr04}
{Carr}, J.~S., {Tokunaga}, A.~T., \& {Najita}, J. 2004, \apj, 603, 213

\bibitem[{{Chen} {et~al.}(2009){Chen}, {Launhardt}, \& {Henning}}]{Chen09}
{Chen}, X., {Launhardt}, R., \& {Henning}, T. 2009, \apj, 691, 1729

\bibitem[{{Cowley} {et~al.}(2012){Cowley}, {Hubrig}, {Castelli}, \&
  {Wolff}}]{Cowley12}
{Cowley}, C.~R., {Hubrig}, S., {Castelli}, F., \& {Wolff}, B. 2012, \aap, 537,
  L6

\bibitem[{{Doppmann} {et~al.}(2011){Doppmann}, {Najita}, {Carr}, \&
  {Graham}}]{Doppmann11}
{Doppmann}, G.~W., {Najita}, J.~R., {Carr}, J.~S., \& {Graham}, J.~R. 2011,
  \apj, 738, 112

\bibitem[{{Fairlamb} {et~al.}(2015){Fairlamb}, {Oudmaijer},
  {Mendigut{\'{\i}}a}, {Ilee}, \& {van den Ancker}}]{Fairlamb15}
{Fairlamb}, J.~R., {Oudmaijer}, R.~D., {Mendigut{\'{\i}}a}, I., {Ilee}, J.~D.,
  \& {van den Ancker}, M.~E. 2015, \mnras, 453, 976

\bibitem[{{Fairlamb} {et~al.}(2017){Fairlamb}, {Oudmaijer}, {Mendigutia},
  {Ilee}, \& {van den Ancker}}]{Fairlamb17}
{Fairlamb}, J.~R., {Oudmaijer}, R.~D., {Mendigutia}, I., {Ilee}, J.~D., \& {van
  den Ancker}, M.~E. 2017, \mnras, 464, 4721

\bibitem[{{Fedele} {et~al.}(2012){Fedele}, {Bruderer}, {van Dishoeck},
  {Herczeg}, {Evans}, {Bouwman}, {Henning}, \& {Green}}]{Fedele12}
{Fedele}, D., {Bruderer}, S., {van Dishoeck}, E.~F., {et~al.} 2012, \aap, 544,
  L9

\bibitem[{{Fedele} {et~al.}(2011){Fedele}, {Pascucci}, {Brittain}, {Kamp},
  {Woitke}, {Williams}, {Dent}, \& {Thi}}]{Fedele11}
{Fedele}, D., {Pascucci}, I., {Brittain}, S., {et~al.} 2011, \apj, 732, 106

\bibitem[{{Fedele} {et~al.}(2008){Fedele}, {van den Ancker}, {Acke}, {van der
  Plas}, {van Boekel}, {Wittkowski}, {Henning}, {Bouwman}, {Meeus}, \&
  {Rafanelli}}]{Fedele08}
{Fedele}, D., {van den Ancker}, M.~E., {Acke}, B., {et~al.} 2008, \aap, 491,
  809

\bibitem[{{Fedele} {et~al.}(2013){Fedele}, {Bruderer}, {van Dishoeck}, {Carr},
  {Herczeg}, {Salyk}, {Evans}, {Bouwman}, {Meeus}, {Henning}, {Green},
  {Najita}, \& {G{\"u}del}}]{Fedele13}
{Fedele}, D., {Bruderer}, S., {van Dishoeck}, E.~F., {et~al.} 2013, \aap, 559,
  A77

\bibitem[{{Folsom} {et~al.}(2012){Folsom}, {Bagnulo}, {Wade}, {Alecian},
  {Landstreet}, {Marsden}, \& {Waite}}]{Folsom12}
{Folsom}, C.~P., {Bagnulo}, S., {Wade}, G.~A., {et~al.} 2012, \mnras, 422, 2072

\bibitem[{{Furlan} {et~al.}(2007){Furlan}, {Sargent}, {Calvet}, {Forrest},
  {D'Alessio}, {Hartmann}, {Watson}, {Green}, {Najita}, \& {Chen}}]{Furlan07}
{Furlan}, E., {Sargent}, B., {Calvet}, N., {et~al.} 2007, \apj, 664, 1176

\bibitem[{{Gaia Collaboration} {et~al.}(2016){Gaia Collaboration}, {Brown},
  {Vallenari}, {Prusti}, {de Bruijne}, {Mignard}, {Drimmel}, {Babusiaux},
  {Bailer-Jones}, {Bastian}, \& et~al.}]{Gaia16}
{Gaia Collaboration}, {Brown}, A.~G.~A., {Vallenari}, A., {et~al.} 2016, \aap,
  595, A2

\bibitem[{{Gaia Collaboration} {et~al.}(2018){Gaia Collaboration}, {Brown},
  {Vallenari}, {Prusti}, {de Bruijne}, {Babusiaux}, {Bailer-Jones}, {Biermann},
  {Evans}, {Eyer}, \& et~al.}]{Gaia18}
---. 2018, \aap, 616, A1

\bibitem[{{Hirota} {et~al.}(2008){Hirota}, {Bushimata}, {Choi}, {Honma},
  {Imai}, {Iwadate}, {Jike}, {Kameya}, {Kamohara}, {Kan-Ya}, {Kawaguchi},
  {Kijima}, {Kobayashi}, {Kuji}, {Kurayama}, {Manabe}, {Miyaji}, {Nagayama},
  {Nakagawa}, {Oh}, {Omodaka}, {Oyama}, {Sakai}, {Sasao}, {Sato}, {Shibata},
  {Tamura}, \& {Yamashita}}]{Hirota08}
{Hirota}, T., {Bushimata}, T., {Choi}, Y.~K., {et~al.} 2008, \pasj, 60, 37

\bibitem[{{Hubrig} {et~al.}(2010){Hubrig}, {Sch{\"o}ller}, {Savanov},
  {Gonz{\'a}lez}, {Cowley}, {Sch{\"u}tz}, {Arlt}, \& {R{\"u}diger}}]{Hubrig10}
{Hubrig}, S., {Sch{\"o}ller}, M., {Savanov}, I., {et~al.} 2010, Astronomische
  Nachrichten, 331, 361

\bibitem[{{Ilee} {et~al.}(2014){Ilee}, {Fairlamb}, {Oudmaijer},
  {Mendigut{\'{\i}}a}, {van den Ancker}, {Kraus}, \& {Wheelwright}}]{Ilee14}
{Ilee}, J.~D., {Fairlamb}, J., {Oudmaijer}, R.~D., {et~al.} 2014, \mnras, 445,
  3723

\bibitem[{{Kama} {et~al.}(2015){Kama}, {Folsom}, \& {Pinilla}}]{Kama15}
{Kama}, M., {Folsom}, C.~P., \& {Pinilla}, P. 2015, \aap, 582, L10

\bibitem[{{K\"aufl} {et~al.}(2004){K\"aufl}, {Ballester}, {Biereichel},
  {Delabre}, {Donaldson}, {Dorn}, {Fedrigo}, {Finger}, {Fischer}, {Franza},
  {Gojak}, {Huster}, {Jung}, {Lizon}, {Mehrgan}, {Meyer}, {Moorwood}, {Pirard},
  {Paufique}, {Pozna}, {Siebenmorgen}, {Silber}, {Stegmeier}, \&
  {Wegerer}}]{Kaufl04}
{K\"aufl}, H.-U., {Ballester}, P., {Biereichel}, P., {et~al.} 2004, in
  \procspie, Vol. 5492, Ground-based Instrumentation for Astronomy, ed.
  A.~F.~M. {Moorwood} \& M.~{Iye}, 1218--1227

\bibitem[{{Kunde} \& {Maguire}(1974)}]{Kunde74}
{Kunde}, V.~R., \& {Maguire}, W.~C. 1974, \jqsrt, 14, 803

\bibitem[{{Mandell} {et~al.}(2012){Mandell}, {Bast}, {van Dishoeck}, {Blake},
  {Salyk}, {Mumma}, \& {Villanueva}}]{Mandell12}
{Mandell}, A.~M., {Bast}, J., {van Dishoeck}, E.~F., {et~al.} 2012, \apj, 747,
  92

\bibitem[{{Mandell} {et~al.}(2008){Mandell}, {Mumma}, {Blake}, {Bonev},
  {Villanueva}, \& {Salyk}}]{Mandell08}
{Mandell}, A.~M., {Mumma}, M.~J., {Blake}, G.~A., {et~al.} 2008, \apjl, 681,
  L25

\bibitem[{{Meeus} {et~al.}(2012){Meeus}, {Montesinos}, {Mendigut{\'{\i}}a},
  {Kamp}, {Thi}, {Eiroa}, {Grady}, {Mathews}, {Sandell}, {Martin-Za{\"i}di},
  {Brittain}, {Dent}, {Howard}, {M{\'e}nard}, {Pinte}, {Roberge},
  {Vandenbussche}, \& {Williams}}]{Meeus12}
{Meeus}, G., {Montesinos}, B., {Mendigut{\'{\i}}a}, I., {et~al.} 2012, \aap,
  544, A78

\bibitem[{{Najita} {et~al.}(1996){Najita}, {Carr}, {Glassgold}, {Shu}, \&
  {Tokunaga}}]{Najita96}
{Najita}, J., {Carr}, J.~S., {Glassgold}, A.~E., {Shu}, F.~H., \& {Tokunaga},
  A.~T. 1996, \apj, 462, 919

\bibitem[{{Najita} \& {{\'A}d{\'a}mkovics}(2017)}]{Najita17}
{Najita}, J.~R., \& {{\'A}d{\'a}mkovics}, M. 2017, \apj, 847, 6

\bibitem[{{Ochsenbein} {et~al.}(2000){Ochsenbein}, {Bauer}, \&
  {Marcout}}]{Ochsenbein00}
{Ochsenbein}, F., {Bauer}, P., \& {Marcout}, J. 2000, \aaps, 143, 23

\bibitem[{{Petrov} {et~al.}(2014){Petrov}, {Kurosawa}, {Romanova}, {Gameiro},
  {Fernandez}, {Babina}, \& {Artemenko}}]{Petrov14}
{Petrov}, P.~P., {Kurosawa}, R., {Romanova}, M.~M., {et~al.} 2014, \mnras, 442,
  3643

\bibitem[{{Pontoppidan} {et~al.}(2010){Pontoppidan}, {Salyk}, {Blake},
  {Meijerink}, {Carr}, \& {Najita}}]{Pontoppidan10}
{Pontoppidan}, K.~M., {Salyk}, C., {Blake}, G.~A., {et~al.} 2010, \apj, 720,
  887

\bibitem[{{Rice} {et~al.}(2006){Rice}, {Armitage}, {Wood}, \&
  {Lodato}}]{Rice06}
{Rice}, W.~K.~M., {Armitage}, P.~J., {Wood}, K., \& {Lodato}, G. 2006, \mnras,
  373, 1619

\bibitem[{{Rothman} {et~al.}(2003){Rothman}, {Barbe}, {Benner}, {Brown},
  {Camy-Peyret}, {Carleer}, {Chance}, {Clerbaux}, {Dana}, {Devi}, {Fayt},
  {Flaud}, {Gamache}, {Goldman}, {Jacquemart}, {Jucks}, {Lafferty}, {Mandin},
  {Massie}, {Nemtchinov}, {Newnham}, {Perrin}, {Rinsland}, {Schroeder},
  {Smith}, {Smith}, {Tang}, {Toth}, {Vander Auwera}, {Varanasi}, \&
  {Yoshino}}]{Rothman03}
{Rothman}, L.~S., {Barbe}, A., {Benner}, D.~C., {et~al.} 2003, \jqsrt, 82, 5

\bibitem[{{Rothman} {et~al.}(2013){Rothman}, {Gordon}, {Babikov}, {Barbe},
  {Chris Benner}, {Bernath}, {Birk}, {Bizzocchi}, {Boudon}, {Brown},
  {Campargue}, {Chance}, {Cohen}, {Coudert}, {Devi}, {Drouin}, {Fayt}, {Flaud},
  {Gamache}, {Harrison}, {Hartmann}, {Hill}, {Hodges}, {Jacquemart}, {Jolly},
  {Lamouroux}, {Le Roy}, {Li}, {Long}, {Lyulin}, {Mackie}, {Massie},
  {Mikhailenko}, {M{\"u}ller}, {Naumenko}, {Nikitin}, {Orphal}, {Perevalov},
  {Perrin}, {Polovtseva}, {Richard}, {Smith}, {Starikova}, {Sung}, {Tashkun},
  {Tennyson}, {Toon}, {Tyuterev}, \& {Wagner}}]{Rothman13}
{Rothman}, L.~S., {Gordon}, I.~E., {Babikov}, Y., {et~al.} 2013, \jqsrt, 130, 4

\bibitem[{{Salyk} {et~al.}(2011{\natexlab{a}}){Salyk}, {Blake}, {Boogert}, \&
  {Brown}}]{Salyk11_2}
{Salyk}, C., {Blake}, G.~A., {Boogert}, A.~C.~A., \& {Brown}, J.~M.
  2011{\natexlab{a}}, \apj, 743, 112

\bibitem[{{Salyk} {et~al.}(2008){Salyk}, {Pontoppidan}, {Blake}, {Lahuis}, {van
  Dishoeck}, \& {Evans}}]{Salyk08}
{Salyk}, C., {Pontoppidan}, K.~M., {Blake}, G.~A., {et~al.} 2008, \apjl, 676,
  L49

\bibitem[{{Salyk} {et~al.}(2011{\natexlab{b}}){Salyk}, {Pontoppidan}, {Blake},
  {Najita}, \& {Carr}}]{Salyk11}
{Salyk}, C., {Pontoppidan}, K.~M., {Blake}, G.~A., {Najita}, J.~R., \& {Carr},
  J.~S. 2011{\natexlab{b}}, \apj, 731, 130

\bibitem[{{Siess} {et~al.}(2000){Siess}, {Dufour}, \& {Forestini}}]{Siess00}
{Siess}, L., {Dufour}, E., \& {Forestini}, M. 2000, \aap, 358, 593

\bibitem[{{Thi} \& {Bik}(2005)}]{Thi05}
{Thi}, W.-F., \& {Bik}, A. 2005, \aap, 438, 557

\bibitem[{{Troutman}(2010)}]{Troutman10}
{Troutman}, M.~R. 2010, PhD thesis, Clemson University

\bibitem[{{Valenti} {et~al.}(2003){Valenti}, {Fallon}, \&
  {Johns-Krull}}]{Valenti03}
{Valenti}, J.~A., {Fallon}, A.~A., \& {Johns-Krull}, C.~M. 2003, \apjs, 147,
  305

\bibitem[{{Valenti} {et~al.}(2000){Valenti}, {Johns-Krull}, \&
  {Linsky}}]{Valenti00}
{Valenti}, J.~A., {Johns-Krull}, C.~M., \& {Linsky}, J.~L. 2000, \apjs, 129,
  399

\bibitem[{{van der Plas} {et~al.}(2008){van der Plas}, {van den Ancker},
  {Fedele}, {Acke}, {Dominik}, {Waters}, \& {Bouwman}}]{vanderPlas08}
{van der Plas}, G., {van den Ancker}, M.~E., {Fedele}, D., {et~al.} 2008, \aap,
  485, 487

\bibitem[{{van der Plas} {et~al.}(2015){van der Plas}, {van den Ancker},
  {Waters}, \& {Dominik}}]{vanderPlas15}
{van der Plas}, G., {van den Ancker}, M.~E., {Waters}, L.~B.~F.~M., \&
  {Dominik}, C. 2015, \aap, 574, A75

\bibitem[{{Walsh} {et~al.}(2015){Walsh}, {Nomura}, \& {van Dishoeck}}]{Walsh15}
{Walsh}, C., {Nomura}, H., \& {van Dishoeck}, E. 2015, \aap, 582, A88

\bibitem[{{Zhu} {et~al.}(2012){Zhu}, {Nelson}, {Dong}, {Espaillat}, \&
  {Hartmann}}]{Zhu12}
{Zhu}, Z., {Nelson}, R.~P., {Dong}, R., {Espaillat}, C., \& {Hartmann}, L.
  2012, \apj, 755, 6

\end{thebibliography}

\end{document}